%% file: SpitzerGalex_V10.tex
\documentclass[preprint]{aastex}	
\usepackage{amsmath}
\usepackage[bookmarks=false]{hyperref}
\usepackage[all]{hypcap}
\usepackage{lscape,longtable}
\usepackage{subfigure}
\usepackage{natbib}
\shorttitle{Star Formation in Brightest Cluster Galaxies}
\shortauthors{Hoffer et al.}

\begin{document}

\title{Infrared and Ultraviolet Star Formation in Brightest Cluster Galaxies in the ACCEPT Sample}

\author{Aaron S. Hoffer}
\affil{Michigan State University, Physics \& Astronomy Dept., East Lansing, MI 48824-2320}
\email{hofferaa@msu.edu}

\author{Megan Donahue}
\affil{Michigan State University, Physics \& Astronomy Dept., East Lansing, MI 48824-2320}
\email{donahue@pa.msu.edu}

\author{Amalia Hicks}
\affil{Michigan State University, Physics \& Astronomy Dept., East Lansing, MI 48824-2320}
\email{hicksam@msu.edu}

\author{R.S. Barthelemy}
\affil{Western Michigan University, Physics Dept., Kalamazoo, MI 49008-5252}
\email{ramon.s.barthelemy@wmich.edu}

\begin{abstract}
We present infrared (IR) and ultraviolet (UV) photometry for a sample of brightest cluster galaxies (BCGs).
The BCGs are from a heterogeneous but uniformly characterized sample, the Archive of {\it Chandra} Cluster Entropy Profile Tables (ACCEPT), of X-ray galaxy clusters 
from the {\it Chandra} X-ray telescope archive with published gas temperature, density, and entropy profiles.
We use archival {\it Galaxy Evolution Explorer} ({\it GALEX}), {\it Spitzer} Space Telescope, and Two Micron All Sky Survey ({\it 2MASS}) observations to assemble spectral
energy distributions (SEDs) and colors for BCGs. We find that while the SEDs of some BCGs follow the expectation of
red, dust-free old stellar populations, many exhibit signatures of recent star formation in the form of
excess UV or mid-IR emission, or both.  We establish a mean near-UV (NUV) to {\it 2MASS} K color of $6.59 \pm 0.34$ for quiescent BCGs. 
We use this mean color to quantify the UV excess associated with star formation in the active BCGs. We use both 
fits to a template of an evolved stellar population and library of starburst models and mid-IR star formation relations
to estimate the obscured star formation rates. 
We show that many of the BCGs in X-ray clusters with low central gas entropy exhibit enhanced 
UV (38\%) and mid-IR emission (43\%) from 8-160 microns, above that expected from an old stellar population. These excesses 
are consistent with on-going star formation activity in the BCG, 
star formation that appears to be enabled by the presence of high density, 
X-ray emitting intergalactic gas in the the core of the cluster of galaxies. 
This hot, X-ray emitting gas may provide the enhanced ambient pressure and some of the fuel to trigger the star formation.
This result is consistent with previous works that showed that BCGs 
in clusters with low central gas entropies host H$\alpha$ emission-line nebulae and radio sources, 
while clusters with high central gas entropy exhibit none of these features. 
{\it GALEX} UV and {\it Spitzer} mid-IR measurements combined provide a complete picture of unobscured and obscured star 
formation occurring in these systems.
We present IR and UV photometry and estimated equivalent continuous 
star formation rates for a sample of brightest cluster galaxies. 

\end{abstract}

\keywords{galaxies: elliptical and lenticular, cD}

\section{Introduction}

The basic story underlying our current models for the formation of galaxies and clusters of galaxies is that baryonic
matter falls into dark matter potential wells, cools to make cold molecular clouds, which then form stars and supermassive black holes.
The state of the gas as it falls, the morphology of the accretion, the source of the dust that catalyzes formation of molecular clouds, the physical
processes determining the gas temperatures and phases are all uncertain. 
Simply put, we do not know the full story of 
how intergalactic gas eventually forms stars and black holes. 

Brightest Cluster Galaxies (BCGs) provide unique opportunities for the 
investigation of the role of hot intergalactic gas in galaxy formation, and in particular its
role in affecting the evolution of the star formation
and active galactic nucleus (AGN) activity in the central galaxy in the most massive dark matter
halos in the universe. 
The intergalactic gas bound to a massive cluster of galaxies -- its intracluster medium (ICM) -- 
outweighs the stars in those galaxies by a factor of 5-10 
\citep[e.g., ][]{1990ApJ...356...32D,1992A&A...254...49A,2007ApJ...666..147G}.
The BCGs in the centers of X-ray clusters where the gas
has a short cooling time (or equivalently, low gas entropy) exhibit signs of activity (e.g. radio sources, emission-line nebulae,
excess blue or ultraviolet light) that are rare in BCGs in other clusters of galaxies 
\citep{1985ApJS...59..447H,1990AJ.....99...14B,2008ApJ...683L.107C,2008ApJ...687..899R,2009MNRAS.398.1698S,2009ApJ...704.1586S}.
The activity in the BCGs of this category of 
clusters has been presented as evidence that hot ICM condenses into cold dusty gas that subsequently
forms stars. Such BCGs may be hosting real-life versions of late-time ($z<1$) 
accretion onto supermassive black holes in central galaxies; but the role of the hot ICM in AGN or star formation activity 
is not entirely clear.

The simplest hypothesis for how hot gas cools when it is confined to a massive dark halo fails.  
The first X-ray observations of the ICM in galaxy clusters indicated that some clusters have a high central gas densities and central 
cooling times shorter than the age of the universe(e.g., \citet{1977MNRAS.180..479F}; \citet{1977ApJ...215..723C}). In this scenario,  
such gas cools slowly, loses pressure support, compresses, allowing gas from the outer parts of the cluster to settle gently into the center.
The inferred mass accretion rates could be as large as $1000~\rm{M_{\sun}~yr^{-1}}$ \citep{1994ARA&A..32..277F}. 
Such clusters were dubbed "cooling flows."  However, higher resolution
X-ray spectroscopy showed that the luminous emission lines one would expect from gas cooling smoothly from $10^8$K to non-X-ray emitting
temperatures were not present \citep{2003ApJ...590..207P}. 
Nevertheless, such clusters do exhibit cool cores with radii $\sim 50-100$ kpc, 
where $kT_{core} \sim 1/2-1/3$ of that found in the outer radii. These clusters are now often called "cool core" clusters.

With spatially resolved X-ray spectroscopy, cool core clusters can be classified by the distribution of gas entropy of the galaxy cluster.  
The cluster entropy is a thermodynamic quantity. Conveniently, in a gas of pure hydrogen emitting thermal bremsstrahlung radiation, the
cooling time can be written down solely in terms of the gas entropy. 
The gas entropy $S$ is proportional to the logarithm of the quanitity $K\rm{=T_X~n_e^{-2/3}}$, conventionally reported in units of keV cm$^{2}$.  
\citet{2006ApJ...643..730D} radially fit entropy profiles with a functional form $\rm{K(r)=K_0+K_x(r/r_x)^{\alpha}}$, where $\rm{K_0}$ is the central entropy in excess above the power law fit. \citet{2008ApJ...683L.107C} extended this procedure to the entire {\it Chandra} archive, creating the Archive of {\it Chandra} Cluster Entropy Profile Tables\footnote[1]{\url{http://www.pa.msu.edu/astro/MC2/accept/}} (ACCEPT).
Galaxy clusters with high central entropy often contain quiescent brightest cluster galaxies (BCGs) or exhibit evidence for significant merger or interactions. 
The empirical boundary between clusters which occasionally host active BCGs and clusters which never host them is \ $\rm{K_0}\sim 30$ $~\rm{keV~cm^2}$, an
entropy associated with an ICM cooling time of $\sim 1$ Gyr \citep{2008ApJ...681L...5V}. Furthermore, about 70\% of the BCGs in
those cool core clusters host radio sources, and about half of those host extended emission-line nebula characteristic of
low-ionization nuclear emission-line regions (LINERs; but are more extended) \citep{1989ApJ...338...48H,1999MNRAS.306..857C,2010ApJ...715..881D}. 
 \citet{2008ApJ...683L.107C} and \citet{2008ApJ...687..899R}
have shown that only those BCGs inhabiting clusters with low central gas entropies (short central gas cooling times, high central gas densities) present 
low-ionization emission-line nebulae (H$\alpha$), blue gradients, or radio sources.

In this paper, we look for signatures strongly associated with star formation, ultraviolet (UV) excesses and 
mid-infrared (mid-IR) emission from dust, in the ACCEPT sample of well-studied X-ray clusters. Since even an evolved stellar population emits some UV (and mid-IR), we characterize the stellar content of the BCG using Two Micron All Sky Survey
({\it 2MASS}) K-band photometry and photometry from the IRAC instrument on the {\it Spitzer} Space Telescope, short-wavelength 3.6 and 4.5 micron bands, where available. 
To estimate the contribution of recent star formation we measure the ultraviolet (UV) emission with the {\it Galaxy Evolution Explorer} ({\it GALEX}) observations. The UV samples 
the peak of emission in short-lived O and B stars, thus tracking recent, unobscured star formation. Most of the star formation in the universe occurs
hidden within cold, dusty molecular clouds. The dust in these clouds absorbs the UV and optical light of buried stars and re-emits this light as
mid-IR thermal radiation typical of dust at $\sim100$ K. Some of this reprocessed emission emerges in the form of features, such as the emission complexes associated with polycyclic aromatic hydrocarbons (PAHs) \citep{2011ApJ...732...40D}. 
Puzzlingly, powerful $\rm{H_2}$ features appear to be nearly ubiquitous in systems with H$\alpha$ nebulae, 
at levels unlikely to be associated with typical
star formation processes \citep{1994ASSL..190..169E,1997MNRAS.284L...1J,2000ApJ...545..670D,2011ApJ...732...40D,2006ApJ...652L..21E}. 
Even colder dust (20-30K) in the far-IR has been seen with {\it Herschel} \citep{2010A&A...518L..46E}\citep{2010A&A...518L..47E}, and \citet{2001MNRAS.328..762E} detected
significant masses of CO. 

The measurements of star formation in BCGs based on UV or mid-IR information to date have been relatively limited.
For example, \citet{2010MNRAS.403..683C} report that star formation efficiency varies little over a wide range of galaxy masses in a massive galaxy sample. 
However, while that sample includes 190 massive galaxies observed with {\it GALEX} and {\it Arecibo}, it has very few BCGs.  
\citet{2010ApJ...715..881D} assessed the UV properties of the BCGs in a representative sample of 30 X-ray selected clusters from the Representative XMM-Newton Cluster Structure Survey (REXCESS) \citep{2007A&A...469..363B}, 
while most UV studies are of a limited set of the most extreme emission-line BCGs \citep[e.g., ][]{2010ApJ...719.1844H,2010ApJ...719.1619O}. 
\citet{2008ApJS..176...39Q} and \citet{2008ApJ...681.1035O} studied 62 BCGs with {\it Spitzer}, selected for their luminous H$\alpha$.
 To expand upon these studies, we present an assessment of the UV, near-IR and mid-IR properties of 
BCGs in a well-studied sample of X-ray clusters. This sample is larger and more diverse than previous studies, as it 
includes quiescent BCGs along with the most extreme cool-core BCGs.  
 In Section 2 we briefly describe the original X-ray cluster sample, and give an overview of the {\it GALEX} and {\it Spitzer} observations. 
 We describe how the BCGs are identified.  
 In Section 3 we discuss the data reduction process for the images in the {\it Spitzer}, {\it GALEX}, and {\it 2MASS} archives. 
 Our discussion and analysis of the data is in Section 4. We present estimates of the equivalent continuous UV and IR star formation rates in this section. UV colors are compared to those in \citet{2010MNRAS.401..433W}. We present a summary of the observations, detections, and emission excesses in Table~\ref{tab:summary}.
 We conclude the paper in Section 5. For all calculations the assumed cosmology is $H_0 = 70 \rm{km~s^{-1}~Mpc^{-1}}$, $\Omega_M=0.3$, $\Omega_{\Lambda}= 0.7$.

\section{Observations}
\subsection{{\it Chandra} X-Ray Observations}
The original galaxy cluster sample is from the ACCEPT database \citep{2009ApJS..182...12C}, which includes 239 galaxy clusters. This sample is a selection of all galaxy clusters in the {\it Chandra} archive as of August 2008 that met a minimum flux criterion. The clusters were selected to construct entropy profiles and provide central entropy estimates. To be able to accurately measure the entropy profiles, temperature gradients were required to have a precision better than $\rm{\Delta kT_X \approx \pm 1.0~keV}$. \citet{2009ApJS..182...12C} therefore required at least three concentric annuli with a minimum of 2500 counts each. The search resulted in 317 observations of 239 galaxy clusters. Six groups from the flux-limited Highest X-ray Flux Galaxy Cluster Sample (HIFLUGCS) sample \citep{2002ApJ...567..716R} were added to the collection and a number of clusters with analysis complications were removed. (All additional objects are listed in \citet{2009ApJS..182...12C}.) This sample is not a formally complete sample, but, by and large, these clusters were not selected to be included in the Chandra program because of the UV and mid-IR properties of their brightest cluster galaxies. An interestingly large fraction of these clusters now have been observed by GALEX and Spitzer, and so the time is right for a uniform analysis of the X-ray, UV, and mid-IR properties of the BCGs in the sample.

\subsection{{\it 2MASS} Observations - BCG identification}
We used the {\it 2MASS} archive and previous literature to determine the locations of the the BCGs in these galaxy clusters (Table~\ref{tab:aor}). 
The brightest cluster galaxies were initially identified by their {\it 2MASS} position. The locations of the BCGs were determined using a visual inspection (including source brightness and morphology) with {\it 2MASS} J-band images 5$\rm{\arcmin\times5\arcmin}$ in size centered on the X-ray centroid to determine the brightest galaxy in the cluster. This visual inspection was followed up with NASA/IPAC Extragalactic Database (NED)\footnote[2]{\url{http://nedwww.ipac.caltech.edu/}} and the Set of Identifications, Measurements, and Bibliography for Astronomical Data (SIMBAD)\footnote[3]{\url{http://simbad.u-strasbg.fr/simbad/}} object searches within 2\arcmin of the X-ray 
centroid to verify the redshifts of the candidate BCGs. 
All objects in the {\it 2MASS} Extended Source Catalog \citep{2003AJ....125..525J} were checked for redshift information and any other indication that they are the brightest galaxy in the cluster. Some BCGs were too distant to have associated {\it 2MASS} catalog entries. 
The BCGs of these distant clusters were identified using a literature search for journal articles indicating the location of the BCG in the cluster, and are named  by their right ascension and declination. For the clusters in the Sloan Digital Sky Survey (SDSS) footprint, color information and brightness in the optical (u'g'r'i'z') from the data release 7 (DR7) were used to verify the BCG selections \citep{2011AAS...21714908L}. In a small number of cases we 
revised the original selection of {\it 2MASS} location (Abell 2034, RXJ1022.1+3830, 4C+55.16, Abell 2069, Abell 368, and Abell 2255). Table~\ref{tab:aor} gives each cluster and the {\it 2MASS} coordinate for the brightest cluster galaxy. 

In Table~\ref{tab:physical} we list the physical separation of the BCGs from the X-ray centroid of their host galaxy clusters. 
While most BCGs lie near the X-ray centroid of their galaxy cluster, consistent with their identification as cD galaxies, there are a few that are very far from the center. 
The physical distance between the X-ray centroid and the BCG is plotted as a histogram in Figure~\ref{fig:centroid}. The BCG we identified is twice as likely to be within 10 projected kpc of its X-ray centroid in low entropy clusters (74\%) compared to high entropy clusters (37\%). Note that all BCGs in a low $\rm{K_0}$ system are within 40 kpc of their cluster's X-ray centroid.

\subsection{{\it GALEX} Observations}
The Galaxy Evolution Explorer ({\it GALEX}) obtains images in the near UV (NUV) at $\lambda _{eff}=2267$ $\rm{\AA}$ (bandpass with a full width at half 
maximun (FWHM) of 269 $\rm{\AA}$) and the far UV (FUV) at $\lambda _{eff}=1516$ $\rm{\AA}$ (FWHM of 616 $\rm{\AA}$) \citep{2005ApJ...619L...1M}. 
There are a total of 168 BCGs in our initial sample with {\it GALEX} observations in the {\it GALEX} archive as of 2011 October. We then 
searched the {\it GALEX} Release Six (GR6) catalog for a UV source within 5$\arcsec~$ of the {\it 2MASS} BCG location. In the cases where there were multiple observations, 
the observation with the highest signal to noise was used. Table~\ref{tab:aor} gives the {\it GALEX} object identifiers for each BCG detected. Note that not all observations will have an object identifier as the BCG may have gone undetected in the {\it GALEX} archive. 

\subsection{{\it Spitzer} Observations}
We analyzed archival {\it Spitzer} Infrared Array Camera (IRAC) and the Multiband Imaging Photometer for SIRTF (MIPS) observations. IRAC has four near infrared 
wavebands at 3.6, 4.5, 5.8, and 8.0 $\mu$m \citep{2004ApJS..154...10F}. MIPS \citep{2004ApJS..154...25R} operates in the mid-IR and has three 
wavebands at 24, 70, and 160 $\mu$m. The {\it Spitzer} imaging observations selected for analysis were aimed within 1$\arcmin$ from the X-ray centroid. The Astronomical Observing Request (AOR) numbers are given in Table~\ref{tab:aor}. There are 79 brightest cluster galaxies in ACCEPT with IRAC observations and 100 ACCEPT BCGs with MIPS observations as of 2010 December.
    
\section{Aperture Photometry and Colors}

\subsection{{\it GALEX} UV Photometry}
We used {\it GALEX} aperture photometry provided in the {\it GALEX} catalog and GALEXView\footnote[4]{\url{http://galex.stsci.edu/GalexView/}}. We chose
apertures to match {\it GALEX} measurements and derive colors with photometry from other 
catalogs (e.g. {\it 2MASS}, SDSS) and with our {\it Spitzer} aperture photometry. 
 The optimal aperture for the UV is determined by comparing the estimate of the total flux given in GALEXView to the circular aperture flux. 
 The circular aperture chosen is the one with the flux measurement nearest to the estimated total flux value.   
 For most BCGs the two largest aperture radii (12.8$\arcsec~$ and 17.3$\arcsec~$) were used. 
 The minimum allowed aperture radius was 9.0$\arcsec~$ to avoid aperture correction (the FWHM of {\it GALEX} observations are $\sim 4.5\arcsec$-6$\arcsec$).  
 The {\it GALEX}-detected UV emission is usually centrally concentrated so generally, the UV emission lies within a radius of 9$\arcsec$ even when the angular size, as seen in the optical, of the galaxy is larger. Therefore, the {\it GALEX} aperture size is an approximate upper limit on the size of the UV star formation region. Some of the UV light is produced by evolved stars \citep[e.g., ][]{1999ARA&A..37..603O} so we use the NUV-K color to estimate how much UV comes from recent star formation. We make photometric measurements within sufficiently large apertures to minimize the degree to which aperture corrections could affect our conclusions. The magnitudes are converted from the magnitudes given in the {\it GALEX} catalog to AB magnitudes using zeropoints of 20.08 magnitudes for the NUV and 18.82 magnitudes for the FUV \citep{2007ApJS..173..682M}. The Galactic extinction corrections are applied from \citet{1998ApJ...500..525S} assuming a ratio of 3.1 for $\rm{A_V} / \rm{E_{B-V}}$. The NUV correction assumed is 3.25$\rm{A_V}$ and the FUV correction is 2.5$\rm{A_V}$. UV photometry is presented in Table~\ref{tab:UVflux}.

\subsection{{\it GALEX} UV Upper Limits}
To estimate the detection threshold for {\it GALEX} observations, we evaluated the cataloged 
fluxes of all the well-detected sources with a magnitude error $<0.35$ ($S/N \gtrsim 3$) within 1$^\circ$ of 
the BCG targets. Our GALEX upper limits are based on detections of peaked
sources, i.e. point sources and compact emission regions. A uniform, extended source that fills the aperture will have a higher
detection threshold than this estimate.  We plot these fluxes as a function of their individual exposure times in Figure~\ref{fig:UVupperlimit}. 
The estimated detection threshold is inferred from the upper envelope of these points, which 
is approximated here by curves $\propto t^{1/2}$. 
For the exposure times typical of the all-sky imaging survey (AIS) 
the  estimate for the upper limit in AB magnitudes is $19+1.25 \times \log{t_{NUV}}$ for an exposure time $t_{NUV}$ in seconds. 
Similarly the function for AB magnitude upper limit for the FUV is $18.5+1.25 \times \log{t_{FUV}}$. This relation underestimates the {\it GALEX}
sensitivity for longer exposure times, longer than $\sim500$ seconds. There are 9 BCGs (Abell 2319, 3C 295, Abell 611, Abell 665, Abell 1942, Abell 2631, CL J1226.9+3332, HCG 62, and Abell 2219) which had UV exposure times greater than 500 seconds and have a nondetection. For these objects we looked in the field and set the upper limit to be equal to the dimmest source that was detected (with a magnitude error less than 0.35). 
We report this estimated 3$\rm{\sigma}$ upper limit for all cases where the BCG was undetected and when the {\it GALEX} source flux had 
a large error ($> 0.35$ mag), indicating a highly uncertain detection. For BCGs with NUV upper limits, the {\it 2MASS} fluxes are matched with a 7$\arcsec$ aperture such that they are similar in size to the GALEX PSF.

\subsection{{\it Spitzer} Near and Mid IR Photometry}
For the vast majority of the observations, {\it Spitzer} photometry was measured from the final pipeline product post-Basic Calibration Data (pbcd). 
The pipeline data were flux calibrated in units of $\rm{MJy~steradian^{-1}}$ \citep{2005PASP..117..978R} from the IRAC pipeline version S18.7.0 and the MIPS pipeline version 16.1.0. For the four IRAC wavebands, fluxes were measured inside a circular aperture with a radius of $r = 14.3 ~\rm{kpc~h_{70}^{-1}}$. 
We wrote an IDL program to perform all aperture flux measurements for {\it Spitzer} \citep{2010ApJ...715..881D}. 
The circular aperture is centered on the BCG location in Table~\ref{tab:aor}. 
The backgrounds were computed from an annulus with an inner radius of 35\arcsec~ and an outer radius of 45\arcsec~ for objects which have an angular radius smaller than 35\arcsec. For objects with flux beyond the nominal aperture, the background was computed with an annulus with an inner radius of 1.1$\times$ the radius for the object and an outer radius 1.3$\times$ the radius of the object. To estimate the mean background counts we fit a Gaussian to a histogram of counts per pixel in the background annulus. This procedure provides a background estimate that is robust to possible sources of contamination (e.g. foreground stars) that increase the counts in a small number of background pixels but do not significantly influence the mean of the Gaussian.

Most of the galaxies detected by MIPS are essentially point sources because the FWHM of the point spread function (PSF) for 24, 70, and 160 $\mu$m are 6\arcsec, 18\arcsec, and 40\arcsec, respectively. We measure MIPS fluxes using the same IDL code. Since not all of the flux from the PSF falls in the aperture, MIPS aperture fluxes are corrected using the same aperture correction methods in \S4.3.4 of the MIPS Handbook\footnote[5]{\url{http://irsa.ipac.caltech.edu/data/SPITZER/docs/mips/mipsinstrumenthandbook/}}. The 24 micron aperture radius is fixed at 13\arcsec~ with a background annulus of 15-25\arcsec~ giving a flux correction factor of 1.167. Similarly, the 70 micron aperture radius is fixed at 35\arcsec~ with a background annulus of 40-60\arcsec, and a correction factor, assuming a 30K source, of 1.22. The 160 micron observations were measured at an aperture radius of 40\arcsec~ with a background annulus of 64-128\arcsec, and a correction factor of 1.752 (also assuming a 30K source). We also derived flux estimates using software provided by the {\it Spitzer Science Center},  APEX in MOPEX \citep{2005PASP..117.1113M}, to 
cross-check our aperture flux measurements. The standard input parameters were used and residual images were created to assess whether the source was completely
subtracted. For all sources with proper subtraction, the flux measurement from APEX was compared to the aperture measurement and we verified they were 
consistent within the cited errors. Only the flux values calculated from apertures are included in Table~\ref{tab:IRflux}. In Table~\ref{tab:summary} detections and excesses are equivalent for the 70 and 160 micron observations as we do not have an {\it a priori} belief that quiescent BCGs should exhibit 70 and 160 micron emission.

For the closest and, likely, spatially-extended BCGs, the fluxes from APEX were systematically lower than the aperture flux estimates.
To determine whether any BCG had extended emission or contamination from unrelated point sources, 
we compared aperture-corrected flux measurements with 13\arcsec~ and 35\arcsec~ apertures, and 
we inspected the 24 micron images for point source contamination within the 35\arcsec~ radius aperture. 
Visible contamination was classified as either insignificant, because the 
difference between the two aperture-corrected estimates was smaller than the statistical uncertainty of those fluxes, 
or significant.  We inspected all detections for possible contamination inside 35\arcsec~ but we only found potential contamination  
in the annulus between the 13\arcsec~ and the 35\arcsec~ radii (i.e. we saw no obvious sources of contamination inside 13\arcsec).
Therefore, we do not expect contamination to affect the 24 micron point source flux measurements listed in Table~\ref{tab:IRflux}. 
However, the existence of any contaminating source seen at 24 microns is flagged for our 
70 and 160 micron photometry in Table~\ref{tab:IRflux}, which uses larger apertures.
(Refer to the footnotes in Table~\ref{tab:IRflux} for a description of the contamination categories.) 

 BCGs which did not have point source contamination visible at 24 microns 
 but showed an increase in flux over that expected for a point source in the larger 35\arcsec~ aperture are considered extended. 
 All the objects which have been identified as such are, unsurprisingly, nearby galaxies. Instead of correcting the fluxes of these objects as if they were 
 point sources at 24 microns, the fluxes for these galaxies are reported for the large apertures we used for the IRAC photometry. 
 (One exception, the BCG NGC 4636, was measured at a 35\arcsec~ radius instead because of significant point source contamination beyond this aperture.)

To more directly account for 70 micron contamination, if a 70 micron source was listed as a detection and the 24 micron measurement indicated contamination, 
the 70 micron image was inspected for contaminating sources. If a 70 micron detected source inside the aperture appears to come from an object other than the BCG 
(i.e. its centroid is consistent with that of a non-BCG galaxy) then the detection was downgraded to a conservative upper limit. 
However, these 70 micron upper limits are based on photometry using a smaller, 16\arcsec~ aperture radius with the corresponding point source 
correction of 1.94 to avoid including flux from extraneous point sources in the upper limit. 
The 70 micron upper limits estimated through this method are noted in the table. 
There are two BCGs, Abell 2744a and MS 04516-0305, that are contaminated at 70 microns as well as 160 microns. 
Upper limits for their 160 micron photometry were found using the same 16\arcsec~ aperture radius with the corresponding point source correction of 4.697.

The standard photometric error of 5\% is used for the IRAC points as the systematic errors were always much larger 
than the statistical errors. For MIPS the standard errors are 10\%, 20\%, 20\% for 24, 70, and 160 microns, respectively. 
These standard errors are usually good estimates except in the case of lower $S/N$ 
detections for which statistical uncertanties are important (i.e. $S/N = 5-20$). We report the total errors (including statistical and systematic uncertainties) 
for MIPS with the flux measurements in Table~\ref{tab:IRflux}.

For MIPS, upper limits were estimated for detections that are below 5$\sigma$. The standard deviation of the observation was calculated in the same manner as \citet{2010ApJ...715..881D}. If the standard aperture flux had a $S/N< 5$ the filtered data were used instead. The background on these data are better controlled, 
but the MIPS Handbook warns that low surface brightness emission in the filtered data will be lost. Therefore, the filtered data were 
only used when the standard source detection fell below the 5$\sigma$ limit. Those filtered images that are still below the 5$\sigma$ detection threshold were assigned a 5$\sigma$ upper limit for that detection.   If a BCG is undetected with the standard mosaic but is detected ($>$5$\sigma$) 
using the filtered data it is considered a filtered detection and is labelled as such in the Table~\ref{tab:IRflux}. There were many observations that were considered filtered detections in our first pass through the data, but were revised to upper limits because of 70 micron contamination from non-BCG sources. 

For a few of the nearest and brightest BCGs there was an issue with the final data products 
in the {\it Spitzer} pipeline. In these cases, the BCG contained a spurious point source 
that was much brighter than the rest of the galaxy. These very bright artifacts proved to not be physical because 
the anomalous levels were not detected in the individual BCD frames. 
We mosaicked the individual BCD frames with the MOPEX software using the standard mosiacking procedure and settings. 
The new mosaic images did not exhibit the spurious point sources. The fluxes were then calculated from the new images and were in agreement 
with the original images if the point source was masked out. Those AORs which required remosaicking are noted in Table~\ref{tab:aor}.  

\subsection{{\it 2MASS} Near IR Observations}
{\it 2MASS} J, H and K fluxes and errors are extracted from the {\it 2MASS} Extended Object Catalog \citep{2003AJ....125..525J}. The catalog provides aperture photometry between 5\arcsec~ and 60\arcsec~ in radius. For a few large galaxies (e.g. M87, M49, NGC 4696) the aperture photometry was taken from the {\it 2MASS} Large Galaxy Atlas. The measurements were converted from the system's Vega magnitudes to Janskys using the AB magnitude conversions (0.9, 1.37, and 1.84 mag for J, H, and K bands, respectively) provided in \citet{2003AJ....126.1090C}. 
We correct {\it 2MASS} magnitudes  for Galactic extinction: $A_K=0.112 A_V$,$A_J = 0.276 A_V$, $A_H = 0.176 A_V$ \citep{1998ApJ...500..525S}. 
In order to derive flux ratios normalized to emission dominated by the old stellar population sampled in the near-infrared, 
we matched apertures in the near-IR with those at other wavelengths. Therefore we estimated {\it 2MASS} 
photometry (presented in Table~\ref{tab:2MASSflux}) for each source in three apertures: (1) the {\it GALEX} aperture for $NUV-K$, (2) the IRAC aperture of $r=14.3 h_{70}^{-1}$ kpc for IRAC to near-IR flux ratios, and (3) the 24 micron aperture (for K-band only). After extinction and k-correction, Figure~\ref{fig:klum} shows that the BCGs have no trend in their K band luminosity (the mean is $1.6^{+0.7}_{-0.4}\times$ $10^{44}$ erg s$^{-1} \rm{h_{70}^{-2}}$) as a function of redshift or $K_0$ of these galaxy clusters. 
\section{Discussion}

\subsection{UV Excess and Color\label{sec:UVcolor}}
The UV excess is determined by comparing $NUV-K$ colors, plotted in Figure~\ref{fig:NUV} against excess entropy $K_0$ from \citet{2009ApJS..182...12C}. 
The baseline for quiescent BCGs is visible in this figure. 
The BCGs with excess UV emission, over and above the UV found in quiescent BCGS, are only in the low $\rm{K_0}$ galaxy clusters in the ACCEPT sample. 
While we find no BCGs with excess UV emission in galaxy clusters with high central entropy, 
there are many quiescent BCGs in low central entropy clusters. 
From our sample we estimate the typical $NUV-K$ color for quiescent BCGs from the mean and standard deviation of all BCGs with 
central entropies above 30 $\rm{keV~cm^2}$. We derive a mean color of inert BCGs is 6.59 $\pm$ 0.34. In contrast, the mean color for BCGs in clusters with central entropies less than 30 $\rm{keV~cm^2}$ is 6.11 $\pm$ 0.99.  We define a 
color excess $\rm{\Delta_c= 6.59 - (NUV - K)}$. This excess will be used in \S~\ref{sec:SFR} to estimate the equivalent continuous star formation rate. The color excess is simply defined such that blue light in excess of quiescent BCGs in high entropy clusters can easily be translated into a UV luminosity associated with continuous unobscured star formation. BCGs are considered to have a NUV excess in Table~\ref{tab:summary} if their NUV-K color is at least 1$\sigma$ bluer than the mean color of inert BCGs. We see that 38\% of low central entropy clusters in our sample have a NUV-K excess. The BCGs with the bluest colors are in Abell 426, Abell 1664, and RX J1504.1-0248 which have colors around 3.0.

We plot the $FUV-NUV$ and $NUV-K$ colors for BCGs in Figure~\ref{fig:FUV}. Contamination from line emission from Ly$\alpha$ may occur 
if the redshifted Ly$\alpha$ line is included in the FUV bandpass (within the FWHM (269 $\rm{\AA}$) of the effective 
wavelength (1516$\rm{\AA}$) of the FUV filter), at redshifts between $0.15-0.36$. The right figure plots only nearby BCGs (z$<0.15$) to address this possible effect. Excluding the BCGs which may be contaminated by line emission (z$>0.15$), we do not detect a significant FUV-NUV color 
difference between bluer BCGs (with NUV-K colors less than 6.3) and redder BCGs (with NUV-K colors greater than 6.3). The mean of the FUV-NUV color for bluer BCGs is $0.73\pm0.57$ while the mean of redder BCGs is $0.79\pm0.30$.

\citet{2010MNRAS.401..433W} uses GALEX and SDSS to measure colors on a sample of 113 nearby (z$<0.1$) optically selected BCGs and compare them to a sample of field galaxies. Also, they compare their results to a sample of 21 X-ray selected BCGs from \citet{2008ApJ...687..899R} which included BCGs in both cool-core and non-cool-core clusters. From Figure 7 in \citet{2010MNRAS.401..433W}, the distribution of the FUV-NUV color is consistent with ours with a mean that better matches the photometry from their outer apertures (radius covers 90\% of the light) than that measured within their inner apertures (radius covers 50\% of the light). Similarly, their NUV-r colors are consistent with our NUV-K colors, after transformation between SDSS r and 2MASS K bands, assuming those bands are only affected by emission from the old stellar population.

\subsection{IR Color}
The ratios of 8.0 to 3.6 micron fluxes track the ratios of infrared emission  from polycyclic aromatic hydrocarbons (PAHs), stochastically heated hot dust grains, and possibly rotationally excited molecular hydrogen and other emission lines  \citep[e.g., ][]{2011ApJ...732...40D} to emission from stars. 
We plot these ratios as a function of redshift in Figure~\ref{fig:8036}. The line shows the expectation for a 
passively evolving stellar population with an age of 10 Gyr at $z=0$. After normalizing the ratio for the stellar population, we determine the total number that are at least 1$\sigma$ above the normalized mean for BCGs in high $K_0$ clusters ($1.014 \pm 0.061$) and refer to those as BCGs with excess 8.0 micron emission in Table~\ref{tab:summary}. 
 The points that lie well above this line are likely to have some form of hot dust and/or PAH emission as the observed IRAC 8.0 micron color is sensitive to only strong PAH features. In Figure~\ref{fig:4536} the IRAC ratio of 4.5 to 3.6 micron fluxes from {\it Spitzer} are plotted against redshift, similarly to the plot from \citet{2008ApJS..176...39Q}. Similar to what we have done for the 8.0 to 3.6 micron ratio, we normalize the 4.5 to 3.6 micron ratio for a passively evolving stellar population with an age of 10 Gyr at $z=0$. We then determine a mean of the normalized ratio for BCGs in high $K_0$ clusters ($1.048 \pm 0.019$). All BCGs at least 1$\sigma$ in excess of the mean are considered to have excess 4.5 micron emission.

For both the 8.0 to 3.6 micron flux ratio and the 4.5 to 3.6 micron flux ratio, the only BCGs with excesses over and above a passively evolving old stellar population are those that inhabit clusters with low central entropies, as shown in 
Figure~\ref{fig:4536} and Figure~\ref{fig:8036}. In Figure~\ref{fig:normalized} the 8.0 to 3.6 micron ratio and the 4.5 to 3.6 micron ratio are strongly correlated ($r=0.92, $15$\sigma$ for objects with mid-IR detections and/or NUV-K excesses), which is expected if the excess 4.5 micron emission is generated by processes related to that producing the 8.0 micron emission. The functional fit plotted is 
\begin{equation}
\rm{log_{10}(F_{8.0\mu m}/F_{3.6\mu m}) = (0.153\pm 0.002) + (5.422\pm 0.021)\times log_{10}(F_{4.5\mu m}/F_{3.6\mu m})}.
\end{equation}
Both the ratios have been normalized for passive evolution. As long as the IRAC calibration was consistent over time, these are precise relative flux ratios, independent of the flux calibration. The absolute flux ratios are precise to about 2\%.  The 8.0 and 4.5 micron bandpasses will include PAH and mid-IR emission line features associated with activity seen in cool core BCGs \citep{2011ApJ...732...40D}. The emission of dust-free, evolved stellar populations in
these same bandpasses is similar to the Rayleigh-Jeans tail of a black body, decreasing steeply to longer wavelengths. We see two BCGs HCG 62 and Abell 1644 that show an excess in both normalized ratios however neither shows a NUV-K excess. Abell 1644 was not observed in MIPS and we expect to see a detection in the 70 micron waveband based on this correlation. HCG 62 has a 70 micron upper limit which may be related to the selection effect that it is a very low $K_0$ galaxy group. 

We assess the presence of a luminous dust component, likely to be obscured star formation but also could be contributed by an AGN, 
by looking at the 24 micron to K-band (2.2 micron) flux ratio plotted against the central entropy in Figure~\ref{fig:24K}. 
We note a similar pattern here as found in the UV excess plots (Figure~\ref{fig:NUV}), that the low $K_0$ galaxy clusters are far more likely to host BCGs with warm dust. The possible exception to this pattern is 
Abell 521, which is a high entropy cluster with an elevated 24 micron to K band flux ratio. However, as seen in \citet{2006A&A...446..417F} there is a low entropy, compact, X-ray corona\citep{2007ApJ...657..197S} (i.e. a ``mini-cooling core'') around the BCG in Abell 521, embedded in a cluster with otherwise high entropy. Excess 24 micron emission is estimated by determining the mean of the 24-K ratio of BCGs in high $K_0$ systems (excluding Abell 521) and any BCG with at least 1$\sigma$ above this mean ($0.063 \pm 0.050$) is considered to have excess 24 micron emission. We see that 43\% of the cool cores in our sample have an excess in their 24 micron to K band ratio. The BCGs with the most extreme 24 micron to K band ratios are in ZwCl 0857.9+2107 and Cygnus A with a ratio of about 20. BCGs in Abell 426 and Abell 1068 also have large ratios around 10. All four objects likely have some AGN contribution. We see the scatter (i.e. standard deviation) in the ratio $\rm{log_{10}(F_{24\mu m}/F_{K})}$ is 0.81 for BCGs in low central entropy clusters.

We can compare IR ratios in BCGs to those of normal star-forming galaxies and starbursts, 
similar to Figure 1 in \citet{2007ApJS..173..377J}. The ratios for the BCGs are plotted in Figure ~\ref{fig:john1} as well as the 
SINGS galaxies \citep{2003PASP..115..928K}. Similar to their sample of a wide range of galaxies, the BCGs in our 
sample have the same colors as star-forming galaxies in SINGS. We note that some of the nearby BCGs have 
a higher ratio of 8 micron to 24 micron emission by a factor of 2.  This ratio may indicate a relatively larger contribution from 
PAH emission over very warm dust.  Also, this bandpass may include contributions from the S(4) transition of molecular hydrogen. Rotationally excited molecular hydrogen lines are extremely luminous in some BCGs, and these same lines are not bright in star forming galaxies (Donahue et al. 2011). It is possible that some of the excess emission at the 8.0 micron may be contributed by molecular hydrogen.
 
\subsection{Star Formation Rates (SFRs) \label{sec:SFR}}
The UV color excess, $\rm{\Delta_c}$ defined in \S~\ref{sec:UVcolor}, can be used to estimate the excess UV luminosity due to unobscured star formation:

\begin{equation}
\rm{L_{SFR}} = \rm{L_{\nu}}(1-10^{-\Delta_c/2.5} ) , 
\end{equation}

where the specfic luminosity $\rm{L_{\nu}}$ is converted from the NUV AB magnitude, corrected for Galactic extinction. The NUV k-correction for a star forming spectrum
is negligible out to moderate redshifts \citep{2010ApJ...719.1844H}. The unobscured star formation rate is then estimated 
from the relation in \citet{1998ARA&A..36..189K} and listed in Table~\ref{tab:sfr}. 
The total UV luminosity is estimated to be $\rm{L_{UV}} \sim \rm{\nu L_{\nu}}$ using $\rm{\nu = c/2267 \AA}$. 
Upper limits are based on 3$\sigma$ uncertainties in UV excesses.

The obscured star formation rate is estimated in two ways, (1)  by fitting \citet{2008ApJS..176..438G} starburst 
models and a model of an old stellar population to the {\it 2MASS} and {\it Spitzer} IRAC/MIPS infrared data points, and (2) from using calibrated conversions
of IR luminosity (mostly 24 and 70 micron luminosities) to SFRs.
In the first case, we present a sum of the two models, with independent normalizations. 
Star formation rates were determined for all BCGs with data from at least {\it 2MASS} and the 24 $\mu$m band of MIPS. 
To estimate rest-frame IR luminosities based on the 24 and 70 micron fluxes, k-corrections were applied such that $\rm{L_{rest}} = k\rm{L_{obs}}$. 
The corrections were found using the best-fit Groves model for that individual galaxy and convolving it with the MIPS bandpass, 
both in the rest frame and the observed frame of the galaxy. The actual 70 micron corrections do not depend very much on the
specific Groves starburst model. However,the 24 micron point usually falls around a minimum in the spectrum, which causes a larger scatter in the relation for a give redshift (up to 30\%)  The 24 micron k-corrections are in the range (0.125-1.056), the 70 micron k-corrections are in the range (0.738-1.879).
The total IR luminosity, $\rm{L_{dust}}$, is estimated by integrating the total scaled starburst model over $\rm{\lambda\lambda}$8-1000$\rm{\mu m}$. 
We plot the UV star formation rate against the IR star formation rate in Figure~\ref{fig:SFR}. 

Calibrated conversions for star formation rates from 24 and 70 micron luminosities were used from \citet{2010ApJ...714.1256C}. 
The 70 micron luminosity conversion to a SFR was from Equations (21) and (22) from this paper, depending on the luminosity of that galaxy. 
The 24 micron SFR relation was from Equation (6) which is from \citet{2005ApJ...632L..79W}. From Figure~\ref{fig:SFRcomp} we 
have a comparison between these conversions and the model calibrated star formation rate. The 24 micron luminosity is not
as good at predicting the bolometric IR luminosity (and the integrated star formation)
because it does not sample as close to the cold dust mid-IR emission peak as the 70 micron luminosity. 
The 70 micron flux is much closer to the peak and is likely a better estimate of the IR luminosity and the obscured 
SFR.

We estimate the IR excess IRX =  $\rm{log_{10}(L_{dust}/L_{UV})}$ and plot it against the $FUV-NUV$ color 
(Figure ~\ref{fig:john6}), 
similar to Figure 6 in the \citet{2007ApJS..173..377J} paper, which presents UV and IR data for a 
sample of star-forming disk galaxies and starburst galaxies. 
In an earlier comparison of BCGs with star-forming galaxies, \citet{2010ApJ...719.1844H}  found that the 
cool core BCGs in their sample tended 
to be bluer in UV color and have a large scatter in IRX compared to those properties in the galaxies in \citet{2007ApJS..173..377J}. We do see the larger scatter in IRX for those BCGs that have a bluer UV color. We note that most of the BCGs in our plot are found in low central entropy clusters because those are the only BCGs with FUV, NUV, and {\it Spitzer} mid-IR detections. 

\subsection{Star Formation and Cluster Entropy Profiles}

We have shown here and in previous works \citep[e.g.,][]{2008ApJ...683L.107C,2009ApJS..182...12C,2008ApJ...687..899R}, that BCGs in clusters with low central entropy ($K_0$) are the only BCGs to exhibit signs of vigorous
star formation. The upper threshold for activity in BCGs appears to be around 30 keV cm$^2$. Table~\ref{tab:summary} presents the subsamples with excess emission. We investigate here to see whether
the strength of the signatures of activity, the UV and mid-IR excess, exhibited any trend with the central entropy floor or other cluster property.

Here we take the 
derived star formation rates as simply indicative of the level of star formation activity. By assuming the star formation is constant, we have taken
a nominal assumption about the conversion factors and the starburst models, and translated luminosities into
SFRs. We are not claiming that the star formation is continuous. Distinctions between continuous star formations, simple single-burst models
of a single age, and convolutions of more complicated star formation histories are well beyond the scope of broad-band photometric data and
global measurements. For example, extremely recent star formation is best tracked with H$\alpha$,
but the H$\alpha$ fluxes available from the literature are typically from long-slit spectra, and therefore can underestimate emission line flux if
some of it is located outside
the central 2\arcsec~ or so. H$\alpha$ can also be affected by dust extinction in heavily obscured regions; H$\alpha$ can be produced 
by mechanisms other than by recombination in star formation regions. 
Mid-IR emission provides a pretty reliable assessment of the obscured star formation energy output, 
since it is like a bolometric measure of luminosity emitted by dust. At low star formation rates, the colder dust, 
heated by evolved stars can contribute to the longer wavelength emission, so the lowest IR SFRs in our sample 
(below about 0.1 solar masses per year) may be regarded as upper limits. The UV light from a galaxy is very sensitive 
to the presence of hot stars if some of their light escapes the galaxy. We do not attempt to correct the UV light for 
internal extinction, so the UV and the mid-IR are sampling complementary components of any star formation-related light. 

A sum of the UV and IR SFRs is therefore a best estimate of something akin to the total star formation rate of the BCG, and even
the most conservative interpretation is that they indicate the current luminosity of star formation in the BCG. 
We do not see any correlation between the entropy profile and the strength of star formation 
signatures (e.g. the UV or the mid-IR luminosities of the BCGs with various X-ray gas quantities, 
$K_0$ or the value of the entropy profile at 20 kpc ($K(r=20~\rm{kpc}$))). In Figure~\ref{fig:k0_SFR} we plot the quantities of SFR and $K_0$. Upon first glance, there may seem to be a trend for the detected lowest entropy systems to have the lowest star-formation luminosities. However, these are  lowest redshift groups in the ACCEPT sample, with lower luminosities and masses overall. They are quite nearby, so the ones that are well-observed by Chandra have entropy profiles that probe the sub kpc-scales. Excluding the groups (or including the upper limits for BCGs in groups without evidence for star formation activity) erases any semblance of a trend. To test that we were not missing a trend because the best fit $K_0$ could be biased high for the more distant clusters (see Cavagnolo et al 2009), we plot $K_0$ and SFR for the BCGs with z between 0.05 and 0.15. In this subsample, no trend is visible.  Furthermore, the expected trend would be that the lowest entropy systems would have the largest star formation luminosities because the gas has the shortest detected cooling times.  Therefore we see no evidence for a simple relation between central gas entropy or cooling time and the estimated SFR.

\subsection{ICM Gas Cooling and Star Formation in BCGs}

While the presence of high density, high pressure intracluster gas seems to be a prerequisite for a BCG to host some star formation, role of the intracluster gas is not quite clear. The current paradigm suggests that some of the hot gas cools and forms stars, but a gas that
has been at X-ray temperatures for some time has likely sputtered away any grains it may have had. The lifetime of a typical Galactic dust
grain in $10^7$ K gas is of order 10 million years \citep{1979ApJ...231...77D}. Dust-free gas forms molecular hydrogen only very slowly \citep[e.g.,][]{2009Natur.459...49B}. 
\citet{2011ApJ...738L..24V} show that for
BCGs with measured reservoirs of CO (and H$_2$), the gas residence time ($= M(H_2) / SFR$ ) for BCGs is very similar to that of star-forming
disk galaxies at SFR$<10 ~\rm{M}_\odot~\rm{yr}^{-1}$, around a Gyr. For BCGs with rapid SFRs, the residence time is similar to that of starbursts
with similar SFRs ($\sim 10^7-10^8$ yrs). They calculate that
if much of the stellar winds and ejecta of evolved stars in the BCG are retained by the BCG, perhaps as a consequence of the higher intracluster
pressures, this gas could fuel much of the existing star formation occuring at a steady rate. Certainly for BCGs with SFR$\sim10$ solar masses per
year or less, the stellar ejecta is a source of material that has mass of similar order of magnitude to any source of cooled ICM gas.

However, for galaxies with gas reservoirs of $10^{10}$ solar masses or more, cooled ICM appears to be required to supply the molecular clouds.
The stellar ejecta or contributions from the ISM of dusty galaxies \citep[e.g.][]{1989ApJ...347L..65S} may provide dusty seeds that may mix with the ICM and significantly
accelerate its cooling.  The larger SFRs cannot be sustained at a steady
rate, given the gas supply, and just as in starburst galaxies, must be a short-term situation. The gas may accumulate over a longer period. Given
that $\sim 1/3$ of low redshift cool core galaxies exhibit H$\alpha$, a similar fraction of cool core BCGs (or possibly fewer, if some of the H$\alpha$ emission
is not related to SF) are in the star-forming state. Therefore, such galaxies could accumulate the ejecta of their stellar inhabitants into
molecular clouds for Gigayears, then experience a burst once a threshold surface density of molecular hydrogen was achieved. 

The empirical correlation between the presence low-entropy ICM and the star formation in the central BCG is incontrovertible. However, the common
interpretation of this correlation that cooled ICM fuels the star formation has not been backed up by a physically plausible theory for  
how the hot ICM cools and makes cold and dusty molecular clouds. The resident stellar population is an obvious source of dust (and cool gas) that
should not be neglected.

\section{Conclusions}
We present photometry for brightest cluster galaxies in the ACCEPT cluster sample, 
derived from {\it GALEX}, {\it Spitzer}, and {\it 2MASS} archival observations. This sample includes 239 clusters which were well-observed by
Chandra up until late 2008, with hot gas entropy profiles uniformly extracted \citep{2009ApJS..182...12C}. We identified the BCGs
in all of the clusters. In our BCG identification, it is twice as likely to be within 10 projected kpc of its X-ray centroid in 
low entropy clusters (74\%) compared to high entropy clusters (37\%).

Similar to what has been seen in other star formation indicators (e.g. H$\alpha$), galaxy clusters with low central gas entropies 
(also known as "cool core" clusters) are the only
clusters to host BCGs with infrared and UV excesses above those from the old stellar population. 
The entropy threshold of 30 keV $\rm{cm^2}$ is consistent with the entropy threshold identified by other work 
\citep{2008ApJ...683L.107C,2008ApJ...687..899R,2009ApJS..182...12C}. 
We found 168 observations by the near UV imaging by {\it GALEX}, of which 112 BCGs were detected. 
We found a mean NUV-K (6.59$\pm$0.34) color seen in quiescent BCGs and use that to quantify excess UV emission in individual BCGs.
Of the 84 clusters with low central gas entropy , 32 (38\%) hosted BCGs with a UV excess, while none of the clusters with high central gas entropy did.
The scatter (i.e. standard deviation) in the NUV-K AB color of BCGs in low entropy clusters is considerably higher at 0.99. 
We did not detect a difference between the mean UV color (FUV-NUV) of BCGs (not including those with possible Ly$\alpha$ contamination), within the error, for low and high entropy clusters. 

Similarly, we detected excess infrared emission in some BCGs in low gas entropy clusters over a large range of infrared bands (e.g. 4.5, 8.0, 24, and 70 microns) and no excess in BCGs in high central entropy clusters. 
The mid-IR emission ratios for BCGs (including quiescent BCGs with mid-IR detections) are consistent with, and span a similar range to, galaxies studied in the Sloan Digital Sky Survey (SDSS) galaxies with a range of star forming properties by \citet{2007ApJS..173..392J}. 
For example, 24 of the obeserved 56 BCGs (43\%) in low entropy clusters show excess 24 micron to K band emission. The 
standard deviation of the ratio $\rm{log_{10}(F_{24\mu m}/F_{K})}$ is 0.81 in these BCGs. 
We also see a strong correlation between excess 4.5 micron and 8.0 micron fluxes that may 
indicate correlated PAH emission in both of these bands, when the PAH emission is strong. 

The excess emission seen in the UV can be used in conjunction with the IR emission to estimate a total star formation rate, accounting for both obscured and unobscured star formation. The UV and IR estimates give complementary information whereas H$\alpha$ may be affected by contaminating contributions from other sources (e.g. dust extinction, shocks) or limited by technique (e.g. incomplete spatial coverage in long slit spectroscopy, contamination by N~{\scshape II} in narrow band imaging). Additionally, the multi-wavelength coverage (as opposed to single band measurements) can help to further constrain possible sources of the excess emission. We see that the near-IR to far-IR emission is consistent with a combination of a starburst model and an old stellar population. Clear signs of these empirical correlations and significant dust emission in some low entropy clusters can help constrain star formation estimates in these BCGs. Aside from the previously noted upper threshold for activity at $K_0 = 30 \rm{keV~cm^2}$, we do not detect a correlation between the level of luminosities or excesses with $K_0$ (or equivalently, central cooling time.) However, whether the gas fueling this activity comes from cooling of the ICM or other processes, is not so clear. A significant, massive evolved stellar population in these galaxies may produce dusty gas which may be confined by the hot gas and it may provide the seeds of condensation for the gas from the hot, and presumably dust-free, intracluster medium \citep{2011ApJ...738L..24V}.

\acknowledgments
Support for this research was provided by  {\it Spitzer} contracts JPL RSA 1377112 (MSU RC065166) and JPL 1353923 (MSU RC065195).
M. Donahue and A. Hicks were partially supported by a Long Term Space Astrophysics grant NASA NNG05GD82G (MSU RC062757). We would also like to thank Deborah Haarsma and Luke Leisman for their helpful discussion on BCG identification and Mark Voit for his comments on the text. This research has made use of the SIMBAD database, operated at CDS, Strasbourg, France. This research has made use of the NASA/IPAC Extragalactic Database (NED) which is operated by the Jet Propulsion Laboratory, California Institute of Technology, under contract with the National Aeronautics and Space Administration.

{\it Facility:} \facility{GALEX}, \facility{Spitzer (IRAC,MIPS)}, \facility{CTIO:2MASS}, \facility{FLWO:2MASS}

\bibliographystyle{apj}	
\bibliography{SpitzerGalex_V10}		

\input{./summarytable.tex}

\input{./IDtable.tex}

\input{./physicaltable.tex}

\input{./UVtable.tex}

\input{./IRtable.tex}

\input{./2MASStable.tex}

\input{./SFRtable.tex}

\input{./SpitzerGalexfigures.tex}

\end{document}

%% file: summarytable.tex
\begin{deluxetable}{ccccc}




\tablecaption{Summary of Observations and Detections\label{tab:summary}}


\tablehead{\colhead{Waveband} & \colhead{Observations} & \colhead{Observations ($K_0 \le 30 \rm{keV~cm^2}$)} & \colhead{Detections} & \colhead{Excess}} 

\startdata
NUV & 168 & 84 & 112 & 32\tablenotemark{a} \\
4.5 micron & 76 & 52 & 76 & 13 \\
8.0 micron & 76 & 52 & 76 & 43 \\
24 micron & 98 & 56 & 94 & 24\tablenotemark{b} \\
70 micron & 65 & 46 & 32 & 32 \\
160 micron & 33 & 21 & 16 & 16 \\
\enddata

\tablenotetext{a}{Defined to be NUV-K color less than 6.25, which is at least one sigma bluer than the mean of the quiescent BCGs.}
\tablenotetext{b}{Defined to be a 24 micron to K band ratio greater than $0.113$, which is at least one sigma greater than the mean of the quiescent BCGs.}
\tablecomments{Refer to appropriate section of text for definitions of Detections and Excess.}

\end{deluxetable}

%% file: IDtable.tex

\clearpage
\begin{landscape}
\begin{deluxetable}{cccccccccccc}
\tabletypesize{\footnotesize}
\setlength{\tabcolsep}{0.01in} 

\tablecaption{Brightest Cluster Galaxy Identifications. \label{tab:aor}}

\tablehead{\colhead{Cluster Name} & \colhead{{\it GALEX} IAU Name} & \colhead{2MASS ID} & \colhead{IRAC AOR} & \colhead{MIPS AOR}}

\startdata

1E0657 56a & \nodata & $06^h58^m16^s.04, -55^\circ56\arcmin36~95\arcsec$ & 12674048, 23089920 & 21981440\tablenotemark{a,b}, 23089664\tablenotemark{a} \\
1E0657 56b & \nodata & $06^h58^m35^s.16, -55^\circ56\arcmin55~8\arcsec$ & 12674048, 23089920 & 21981440\tablenotemark{a,b}, 23089664\tablenotemark{a} \\
2A 0335+096 & 033840.6+095812\tablenotemark{c} & 03384056+0958119 & 18646528 & 18636544\tablenotemark{a,b} \\
2PIGG J0011.5-2850 & 001121.5-285114\tablenotemark{d}, 001121.5-285115\tablenotemark{c} & 00112166-2851158 & \nodata & \nodata \\
2PIGG J2227.0-3041 & 222754.3-303433\tablenotemark{e} & 22275446-3034321 & \nodata & \nodata \\
3C 28.0 & 005550.4+262438\tablenotemark{e} & 00555058+2624366 & 18648320 & 18648064\tablenotemark{a,b} \\
3C 295 & 141120.4+521210\tablenotemark{c} & 14112046+5212095 & \nodata & \nodata \\
3C 388 & 184402.5+453329\tablenotemark{e} & 18440241+4533298 & 10925056 & 4710144\tablenotemark{a,b,f}, 21693696\tablenotemark{a} \\
\enddata

\tablenotetext{a}{MIPS observation includes a 24$\mu$m observation.}
\tablenotetext{b}{MIPS observation includes a 70$\mu$m observation.}
\tablenotetext{c}{{\it GALEX} observation only includes FUV measurement.}
\tablenotetext{d}{{\it GALEX} observation only includes NUV measurement.}
\tablenotetext{e}{{\it GALEX} observation includes both NUV and FUV measurement.}
\tablenotetext{f}{MIPS observation includes a 160$\mu$m observation.}
\tablenotetext{g}{{\it Spitzer} observations were made by Donahue as part of the DDT program 488.}
\tablenotetext{h}{Due to an anomalous point source in the pbcd reduction, the images needed to be reduced from the bcd frames.}
\tablenotetext{i}{{\it GALEX} observation was taken as a Guest Investigator for \citet{2010ApJ...719.1844H}.}
\tablenotetext{j}{IRAC observation only has bands 1 and 3 on target.}
\tablenotetext{k}{IRAC observation only has bands 2 and 4 on target.}

\end{deluxetable}
\end{landscape}

%% file: physicaltable.tex
\begin{deluxetable}{ccccccccc}

\tabletypesize{\footnotesize}
\setlength{\tabcolsep}{0.01in} 
\tablecaption{Physical Properties. \label{tab:physical}}

\tablehead{\colhead{Cluster Name} & \colhead{Redshift} & \colhead{Central Entropy\tablenotemark{a}} & \colhead{$K_{100}$\tablenotemark{a}} & \colhead{alpha\tablenotemark{a}} & \colhead{Scale} & \colhead{IR Radius} & \colhead{UV Radius} & \colhead{Centroid Offset\tablenotemark{a}}\\ 
\colhead{} & \colhead{} & \colhead{($\rm{keV~cm^2}$)} & \colhead{($\rm{keV~cm^2}$)} & \colhead{} & \colhead{(kpc/\arcsec)} & \colhead{(\arcsec)} & \colhead{(\arcsec)} &\colhead{(kpc)} } 

\startdata
1E0657 56a & 0.2960 & 307.45 & 18.55 & 1.88 & 4.41 & 3.24 & \nodata & 503.92 \\
1E0657 56b & 0.2960 & 307.45 & 18.55 & 1.88 & 4.41 & 3.24 & \nodata & 216.72 \\
2A 0335+096 & 0.0347 & 7.14 & 138.64 & 1.52 & 0.69 & 20.7 & 12.8 & 1.00 \\
2PIGG J0011.5-2850 & 0.0753 & 101.98 & 214.68 & 0.84 & 1.43 & 10.01 & 17.3 & 2.10 \\
2PIGG J2227.0-3041 & 0.0729 & 17.13 & 113.95 & 1.37 & 1.39 & 10.31 & 17.3 & 0.59 \\
3C 28.0 & 0.1952 & 23.85 & 107.82 & 1.79 & 3.24 & 4.42 & 12.8 & 3.55 \\
3C 295 & 0.4641 & 14.54 & 81.95 & 1.47 & 5.86 & 2.44 & 9 & 16.98 \\
3C 388 & 0.0917 & 17.03 & 214.3 & 0.76 & 1.71 & 8.38 & 17.3 & 1.44 \\
\enddata

\tablenotetext{a}{Quantities defined in \citet{2009ApJS..182...12C} are from radially fit entropy profiles with a functional form $\rm{K(r)=K_0+K_x(r/r_x)^{\alpha}}$, where $\rm{K_0}$ is the central entropy in excess above the power law fit, $\rm{K_{100}}$ is the entropy profile normalization at 100$h_{70}^{-1}$ kpc from the X-ray centroid, and alpha is the best-fit power law index.}
\tablenotetext{b}{UV data is taken from \citet{2010ApJ...719.1844H}}
\tablecomments{Redshift information and central entropies are reproduced from the ACCEPT database. The size scale is calculated as the angular distance size assuming the standard cosmology in the paper. The IR radii are set at 14.3$h_{70}^{-1}$ kpc in size and are used for {\it 2MASS} and IRAC aperture measurements except in the case where the aperture is below 5$\arcsec$. In this case the aperture is set to be 5$\arcsec$ to minimize large aperture corrections. The UV radius is set by the aperture photometry in the {\it GALEX} database which most closely matches the GALEXView total flux measurement. We compare the BCG positions that we derive from the {\it 2MASS} locations in Table~\ref{tab:aor} and compare that to the X-ray centroid in the ACCEPT database. These distances are presented in a histogram in Figure~\ref{fig:centroid}}.

\end{deluxetable}

%% file: UVtable.tex
\begin{deluxetable}{ccccccccccccccc}
\tabletypesize{\scriptsize}
\setlength{\tabcolsep}{0.04in} 
\tablecaption{Fluxes Matched to UV Aperture.\label{tab:UVflux}}

\tablehead{\colhead{Name} & \colhead{Flux} & \colhead{Error} &\colhead{Flux} & \colhead{Error} &\colhead{Flux} & \colhead{Error} &\colhead{Flux} & \colhead{Error} &\colhead{Flux} & \colhead{Error}\\
\colhead{Cluster Name} & \colhead{NUV} & \colhead{NUV} & \colhead{FUV} & \colhead{FUV} & \colhead{J} & \colhead{J} & \colhead{H} & \colhead{H} & \colhead{K} & \colhead{K} \\ 
\colhead{} & \colhead{(mag)} & \colhead{(mag)} & \colhead{(mag)} & \colhead{(mag)} & \colhead{(mag)} & \colhead{(mag)} & \colhead{(mag)} & \colhead{(mag)} & \colhead{(mag)} & \colhead{(mag)}} 

\startdata

1E0657 56a & \nodata & \nodata & \nodata & \nodata & \nodata & \nodata & \nodata & \nodata & \nodata & \nodata \\
1E0657 56b & \nodata & \nodata & \nodata & \nodata & \nodata & \nodata & \nodata & \nodata & \nodata & \nodata \\
2A 0335+096 & 18.05 & 0 & 18.50 & 0 & 13.48 & 0.0130 & 13.13 & 0.0160 & 13.09 & 0.0160 \\
2PIGG J0011.5-2850 & 20.07 & 0.1093 & 20.65 & 0.2108 & 13.15 & 0.0080 & 12.90 & 0.0100 & 13.05 & 0.0131 \\
2PIGG J2227.0-3041 & 18.40 & 0.0176 & 19.85 & 0.0568 & 12.98 & 0.0070 & 12.75 & 0.0090 & 12.90 & 0.0106 \\
3C 28.0 & 20.51 & 0.3453 & 20.56 & 0.2914 & 15.54 & 0.0660 & 15.35 & 0.0720 & 15.11 & 0.0697 \\
3C 295 & 23.17 & 0 & 23.80 & 0.3245 & 16.38 & 0.1570 & 16.18 & 0.2230 & 15.76 & 0.1470 \\
3C 388 & 19.95 & 0.0507 & 20.97 & 0.1878 & 13.29 & 0.0100 & 13.10 & 0.0120 & 13.28 & 0.0139 \\
\enddata
\tablenotetext{a}{BCGs are also in the \citet{2010ApJ...719.1844H} sample.}
\tablenotetext{b}{BCG {\it GALEX} fluxes are taken from \citet{2010ApJ...719.1844H}.}
\tablecomments{Fluxes reported with errors equal to 0 are 3$\sigma$ upper limits. For NUV upper limits, the {\it 2MASS} flux is matched with a 7$\arcsec$ aperture such that it is similar in size to the GALEX PSF.}

\end{deluxetable}

%% file: IRtable.tex
\begin{landscape}
\begin{deluxetable}{ccccccccccccccc}
\setlength{\tabcolsep}{0.01in} 
\tabletypesize{\footnotesize}

\tablecaption{{\it Spitzer} Aperture Flux.\label{tab:IRflux}}

\tablehead{\colhead{Name} & \colhead{Flux} & \colhead{Error}& \colhead{Flux} & \colhead{Error}& \colhead{Flux} & \colhead{Error}& \colhead{Flux} & \colhead{Error}& \colhead{Flux} & \colhead{Error}& \colhead{Flux} & \colhead{Error}& \colhead{Flux} & \colhead{Error}\\
\colhead{Cluster Name} & \colhead{3.6$\mu m$} & \colhead{3.6$\mu m$} & \colhead{4.5$\mu m$} & \colhead{4.5$\mu m$} & \colhead{5.8$\mu m$} & \colhead{5.8$\mu m$} & \colhead{8.0$\mu m$} &\colhead{8.0$\mu m$} & \colhead{24$\mu m$} & \colhead{24$\mu m$} & \colhead{70$\mu m$} & \colhead{70$\mu m$} & \colhead{160$\mu m$}& \colhead{160$\mu m$}\\ 
\colhead{} & \colhead{(mJy)} & \colhead{(mJy)} & \colhead{(mJy)} & \colhead{(mJy)} & \colhead{(mJy)} & \colhead{(mJy)} & \colhead{(mJy)} & \colhead{(mJy)}  & \colhead{(mJy)} & \colhead{(mJy)}& \colhead{(mJy)} & \colhead{(mJy)} & \colhead{(mJy)} & \colhead{(mJy)} }

\startdata
1E0657 56a    & 0.47 & 0.02 & 0.38 & 0.02 & 0.22 & 0.01 & 0.14 & 0.01 & 0.15 & 0.02 & 5.29 &  0 & \nodata & \nodata \\
1E0657 56b    & 0.57 & 0.03 & 0.46 & 0.02 & 0.26 & 0.01 & 0.17 & 0.01 & 0.11 & 0.02 & 1.43\tablenotemark{a} &  0 & \nodata & \nodata \\
2A 0335+096    & 26.12 & 1.31 & 15.93 & 0.80 & 14.79 & 0.74 & 9.94 & 0.50 & 2.40\tablenotemark{b} & 0.24 & 77.10 & 20.36 & \nodata & \nodata \\
2PIGG J0011.5-2850    & \nodata & \nodata & \nodata & \nodata & \nodata & \nodata & \nodata & \nodata & \nodata & \nodata & \nodata & \nodata & \nodata & \nodata \\
2PIGG J2227.0-3041    & \nodata & \nodata & \nodata & \nodata & \nodata & \nodata & \nodata & \nodata & \nodata & \nodata & \nodata & \nodata & \nodata & \nodata \\
3C 28.0    & 0.88 & 0.04 & 0.67 & 0.03 & 0.35 & 0.02 & 0.43 & 0.02 & 0.55\tablenotemark{c} & 0.08 & 6.78 &  0 & \nodata & \nodata \\
3C 295    & \nodata & \nodata & \nodata & \nodata & \nodata & \nodata & \nodata & \nodata & \nodata & \nodata & \nodata & \nodata & \nodata & \nodata \\
3C 388    & 6.20 & 0.31 & 4.11 & 0.21 & 2.70 & 0.13 & 1.88 & 0.09 & 2.15 & 0.22 & 18.79\tablenotemark{d} & 4.16 & 7.13 &  0 \\
\enddata
\tablenotetext{a}{Originally, aperture flux measurement indicated a detection. However, 24 micron flux measurement and visual inspection indicated contamination in the aperture where flux is likely from an unrelated source. The reported measurement is now an upper limit computed using the point source estimate at 16\arcsec.}
\tablenotetext{b}{Source not extended but has significant contamination. Point source measurement at 35\arcsec~ radius is greater than 10 percent error margin.} 
\tablenotetext{c}{Source not extended but has mild source contamination. Point source measurement at 35\arcsec~ radius is within 10 percent error.} 
\tablenotetext{d}{Flux measurement is a filtered detection.}
\tablenotetext{e}{IRAC flux measurements derived from a BCD image that was remosaicked.}
\tablenotetext{f}{Source is extended in the MIPS 24 micron image. 24 micron flux measured within the aperture (see Table~\ref{tab:physical}). Note an aperture radius of 35\arcsec~ was used for NGC4636 because of significant point source contamination outside of 35\arcsec.} 
\tablecomments{Fluxes reported with errors equal to 0 are 5$\sigma$ upper limits.}
\end{deluxetable}

\end{landscape}

%% file: 2MASStable.tex
\clearpage
\begin{deluxetable}{ccccccccc}

\tabletypesize{\scriptsize}

\tablecaption{{\it 2MASS} Aperture Flux.\label{tab:2MASSflux}}

\tablehead{\colhead{Name} & \colhead{Flux} & \colhead{Error}& \colhead{Flux} & \colhead{Error}& \colhead{Flux} & \colhead{Error}& \colhead{Flux} & \colhead{Error}\\
\colhead{Cluster Name} & \colhead{J} & \colhead{J} & \colhead{H} &\colhead{H} & \colhead{K} &\colhead{K} &\colhead{K24} &\colhead{K24}\\
\colhead{} & \colhead{(mJy)} & \colhead{(mJy)} & \colhead{(mJy)} & \colhead{(mJy)} & \colhead{(mJy)} & \colhead{(mJy)} & \colhead{(mJy)} &\colhead{(mJy)}}

\startdata
1E0657 56a & \nodata & \nodata & \nodata & \nodata & \nodata & \nodata & \nodata & \nodata \\
1E0657 56b & \nodata & \nodata & \nodata & \nodata & \nodata & \nodata & \nodata & \nodata \\
2A 0335+096 & 65.69 & 0.54 & 72.49 & 0.94 & 57.94 & 0.83 & 40.23 & 0.50 \\
2PIGG J0011.5-2850 & 12.83 & 0.22 & 15.56 & 0.34 & 14.53 & 0.41 & \nodata & \nodata \\
2PIGG J2227.0-3041 & 15.75 & 0.26 & 19.44 & 0.37 & 17.22 & 0.41 & \nodata & \nodata \\
3C 28.0 & 1.26 & 0.12 & 1.64 & 0.17 & 1.68 & 0.19 & 3.35 & 0.51 \\
3C 295 & 0.76 & 0.11 & 1.03 & 0.18 & 1.31 & 0.17 & \nodata & \nodata \\
3C 388 & 10.31 & 0.18 & 12.48 & 0.28 & 11.57 & 0.32 & 15.38 & 0.49 \\
\enddata

\end{deluxetable}

%% file: SFRtable.tex
\begin{landscape}
\begin{deluxetable}{ccccccccc}
\setlength{\tabcolsep}{0.01in} 
\tabletypesize{\footnotesize}

\tablecaption{Star Formation Rates:\label{tab:sfr}}

\tablehead{\colhead{Name} & \colhead{SFR} & \colhead{Error} & \colhead{SFR} & \colhead{SFR} & \colhead{Error} & \colhead{SFR} & \colhead{Error} & \colhead{Mass} \\ 
\colhead{Cluster Name} & \colhead{UV} & \colhead{UV} & \colhead{IR} & \colhead{70$\mu m$} & \colhead{70$\mu m$} & \colhead{24$\mu m$} & \colhead{24$\mu m$} & \colhead{Stellar Mass} \\ 
\colhead{} & \colhead{($M_{\sun}~yr^{-1}$)}& \colhead{($M_{\sun}~yr^{-1}$)} & \colhead{($M_{\sun}~yr^{-1}$)} & \colhead{($M_{\sun}~yr^{-1}$)} & \colhead{($M_{\sun}~yr^{-1}$)} & \colhead{($M_{\sun}~yr^{-1}$)} & \colhead{($M_{\sun}~yr^{-1}$)} & \colhead{($10^{10} M_{\sun}$)}} 

\startdata
1E0657 56a & \nodata & \nodata & 2.12 & 5.22 & 0 & 0.91 & 0.11 & 5.52 \\
1E0657 56b & \nodata & \nodata & 1.64 & 1.65 & 0 & 0.76 & 0.12 & 6.68 \\
2A 0335+096 & \nodata & \nodata & 0.72 & 0.94 & 0.25 & 0.21 & 0.02 & 5.58 \\
2PIGG J0011.5-2850 & 0.92 & 0 & \nodata & \nodata & \nodata & \nodata & \nodata & \nodata \\
2PIGG J2227.0-3041 & 1.82 & 0.33 & \nodata & \nodata & \nodata & \nodata & \nodata & \nodata \\
3C 28.0 & 2.30 & 0.37 & 13.51 & 2.15 & 0 & 2.05 & 0.29 & 4.40 \\
3C 295 & \nodata & \nodata & \nodata & \nodata & \nodata & \nodata & \nodata & \nodata \\
3C 388 & 1.14 & 0 & 1.78 & 1.61 & 0.36 & 1.47 & 0.15 & 7.06 \\
\enddata

\tablecomments{A star formation rate uncertainty of 0 identifies the quoted rate as a 3$\sigma$ upper limit.}
\tablecomments{UV SFR are calculated for all objects, including those in high $\rm{K_0}$ systems. Upper limits are then calculated using the uncertainties on the NUV, K, and inert BCG color. If a star formation rate is consistent with a SFR of zero within 3$\sigma$, a 3$\sigma$ upper limit is reported.}
\tablecomments{The IR SFR is estimated by a fit to the \citet{2008ApJS..176..438G} models. The 24 and 70 micron SFRs are estimated from empirical relationships reported in \citet{2010ApJ...714.1256C} and the uncertainties are calculated using the 24 and 70 micron flux uncertainties, respectively.}

\end{deluxetable}
\end{landscape}

%% file: SpitzerGalexfigures.tex
\begin{figure}
\includegraphics[scale = .8]{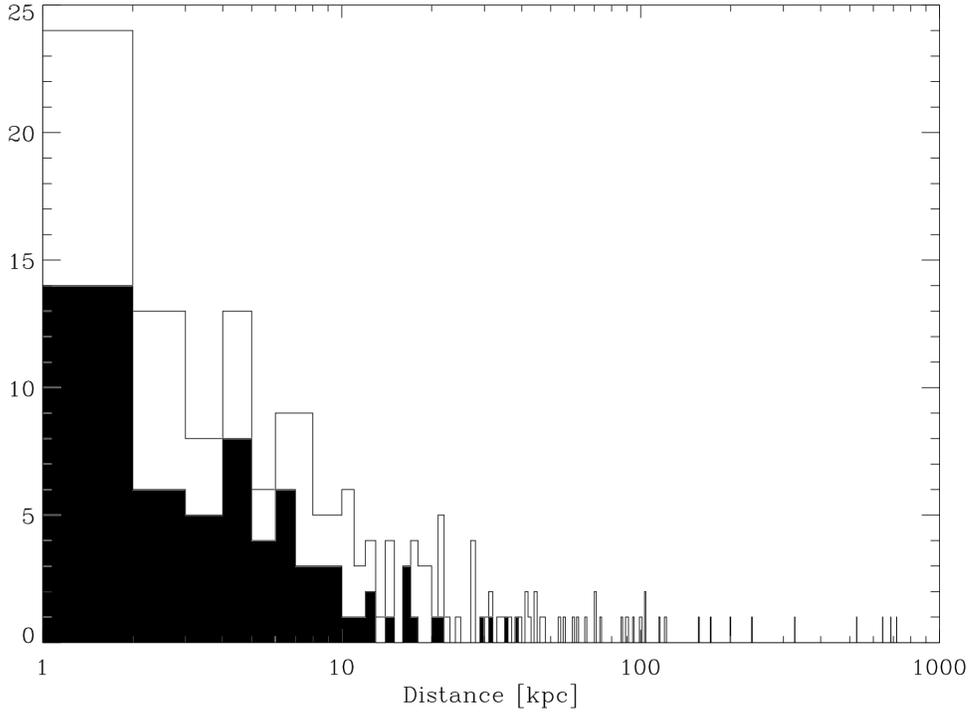}
\caption{The projected physical distance, in kpc h$_{70}^{-1}$, between X-ray centroid and the BCG we identified. All BCGs with an RA and Dec, not just those with {\it GALEX} and {\it Spitzer} data are plotted here. The the shaded region highlights BCGs in low $\rm{K_0}$ clusters. In high central entropy systems, 37\% of BCGs lie within 10 kpc of the X-ray centroid, while the percentage is increased to 74\% for low central entropy systems. All BCGs which lie greater than 40 kpc away from their X-ray centroid are in high $\rm{K_0}$ clusters.}
\label{fig:centroid}
\end{figure}

\begin{figure}
\includegraphics[scale = .8]{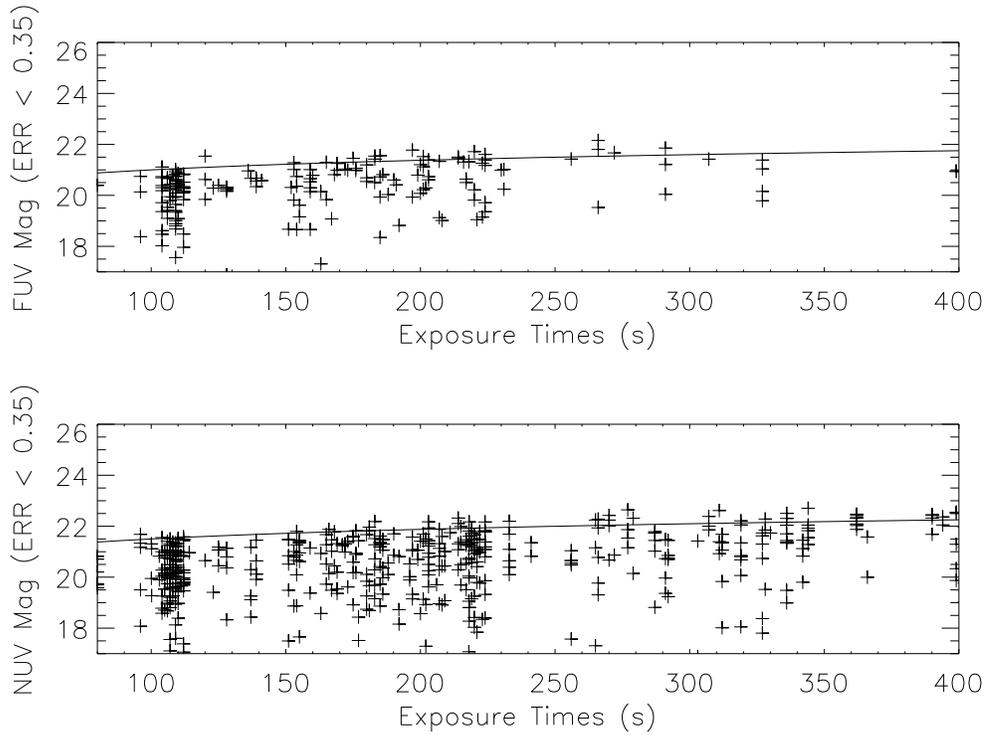}
\caption{The UV magnitudes (AB scale) for all UV sources within 1\arcmin~ of the BCG locations (regardless of identity) 
with flux uncertainties less than 0.35 magnitudes. The upper envelope of this
distribution serves as a basis for estimating the upper limit fluxes for undetected or poorly-detected BCGs for exposure times less than 400 seconds:
$FUV_{UL} = 18.5 + 1.25 \log_{10}{t}$ and $NUV_{UL} = 19.0 + 1.25 \log_{10}{t}$. We considered all {\it GALEX} detections with magnitude errors $> 0.35$ to be poorly detected.}
\label{fig:UVupperlimit}
\end{figure}

\begin{figure}
\includegraphics[scale = .8]{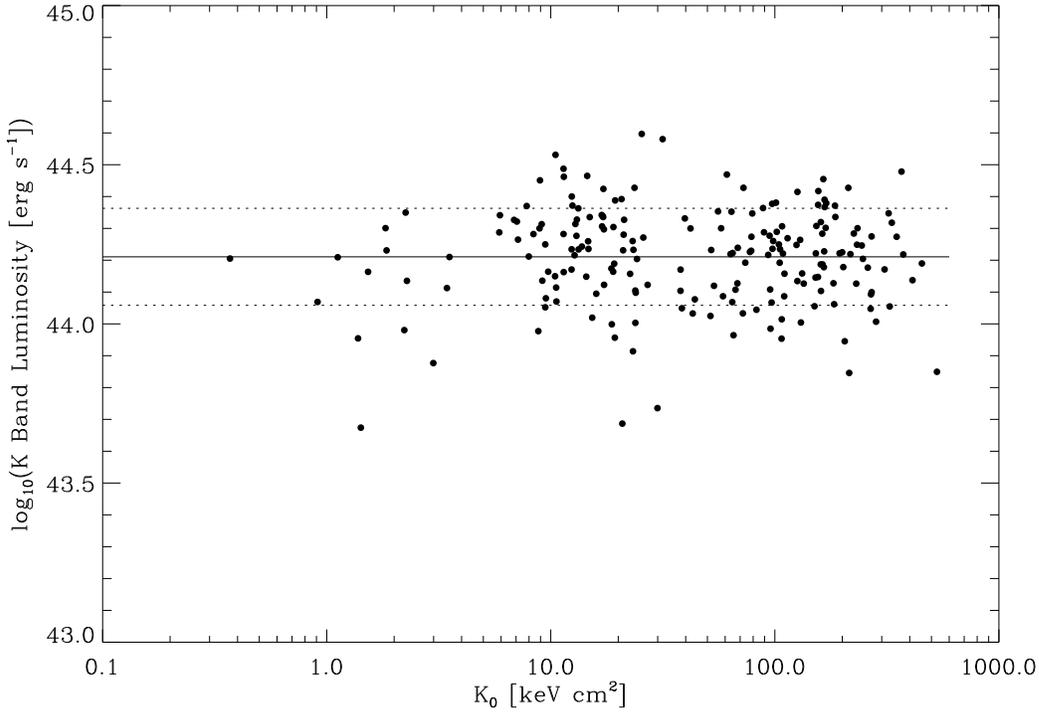}
\caption{The logarithm of K band luminosity and central entropy of the cluster. The K band luminosity is calculated from the flux inside $14.3 ~\rm{kpc~h_{70}^{-1}}$ kpc radius. The luminosities are k-corrected assuming passive evolution. The solid horizontal line represents the mean (1.6$\times$ $10^{44}$ erg s$^{-1} \rm{h_{70}^{-2}}$ ) of the data points while the dotted lines are the 1$\sigma$ error (+0.7$\times$ $10^{44}$ erg s$^{-1} \rm{h_{70}^{-2}}$, -0.4$\times$ $10^{44}$ erg s$^{-1} \rm{h_{70}^{-2}}$ ) on the mean.}
\label{fig:klum}
\end{figure}

\begin{figure}
\includegraphics[scale = .8]{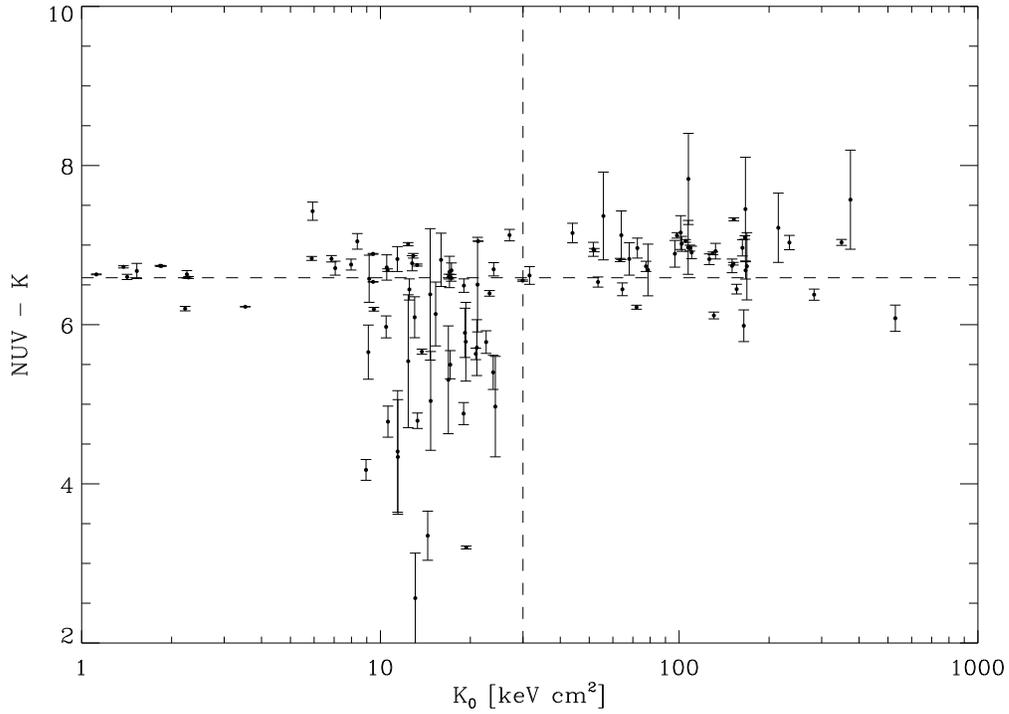}
\caption{NUV-K as a function of cluster central entropy. The vertical dashed line is at 30~$\rm{keV~cm^2}$, our cutoff for the definition of low entropy clusters. Note the large color distribution for low $\rm{K_0}$ objects, while the high entropy objects have a more consistent redder color.
The K band fluxes have been k-corrected assuming passive evolution. The horizontal dashed line represents a NUV-K color of 6.59 magnitudes, the mean of the BCGs in clusters with with $\rm{K_0} \le 30~\rm{keV~cm^2}$. There may appear to be a trend with the low entropy objects, but this is a selection effect where the lowest entropy objects that are observed are also the nearest objects.}
\label{fig:NUV}
\end{figure}

\begin{figure}
\centering
\mbox{\subfigure{\includegraphics[scale = .4]{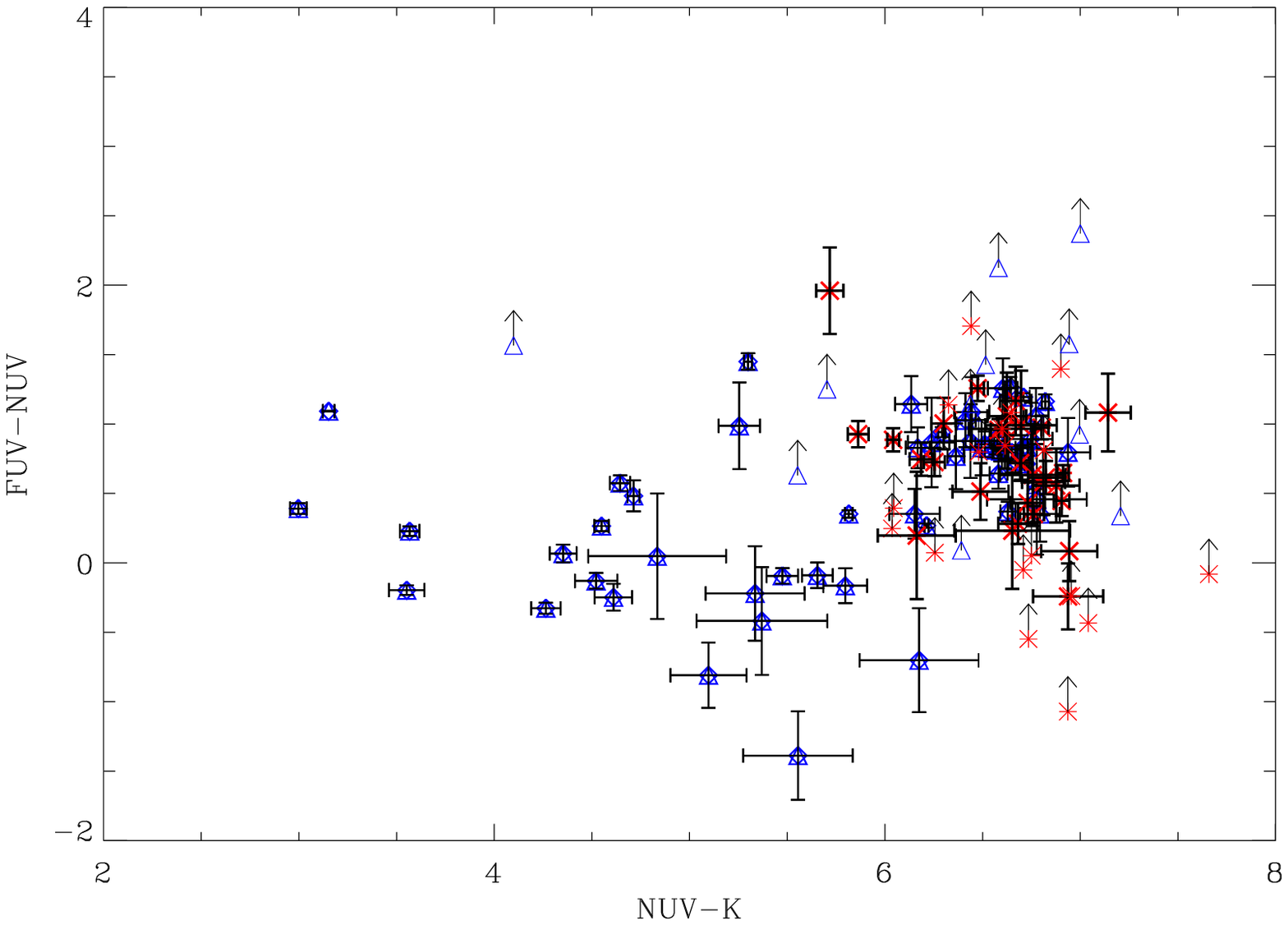}}\quad
\subfigure{\includegraphics[scale=0.4]{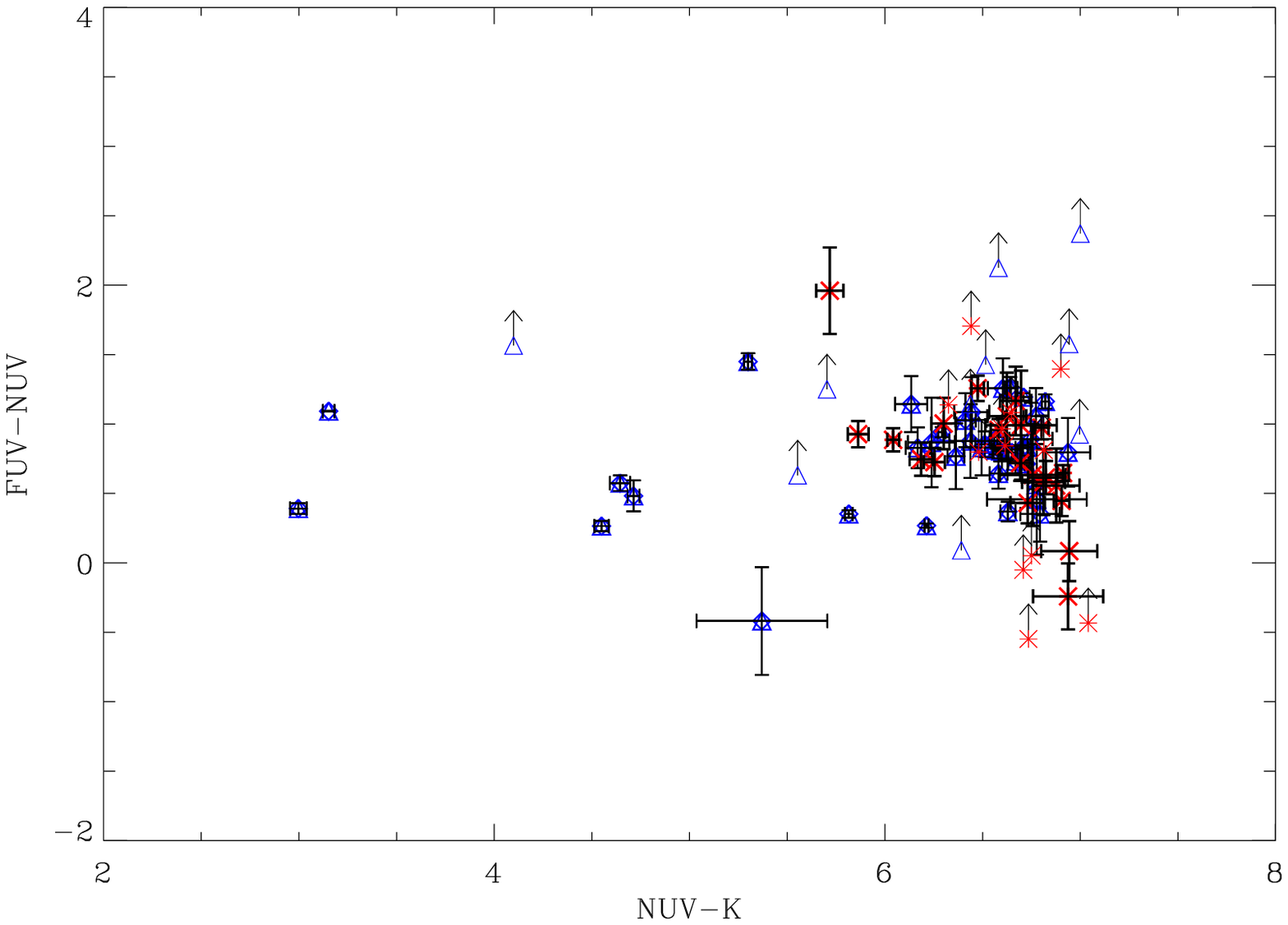}} }
\caption{FUV-NUV and NUV-K colors. The blue triangles are BCGs in low $\rm{K_0}$ clusters ($\le \rm{30}~\rm{keV~cm^2}$) while the red asterisks are BCGs in high entropy clusters. The left plot includes all of the BCGs while the right plot only includes nearby (z$<0.15$) BCGs demonstrating that the bluest FUV-NUV colors, in the left hand plot, are likely arising because of contributions from Ly$\alpha$.}
\label{fig:FUV}
\end{figure}

\begin{figure}
\centering
\mbox{\subfigure{\includegraphics[scale = .4]{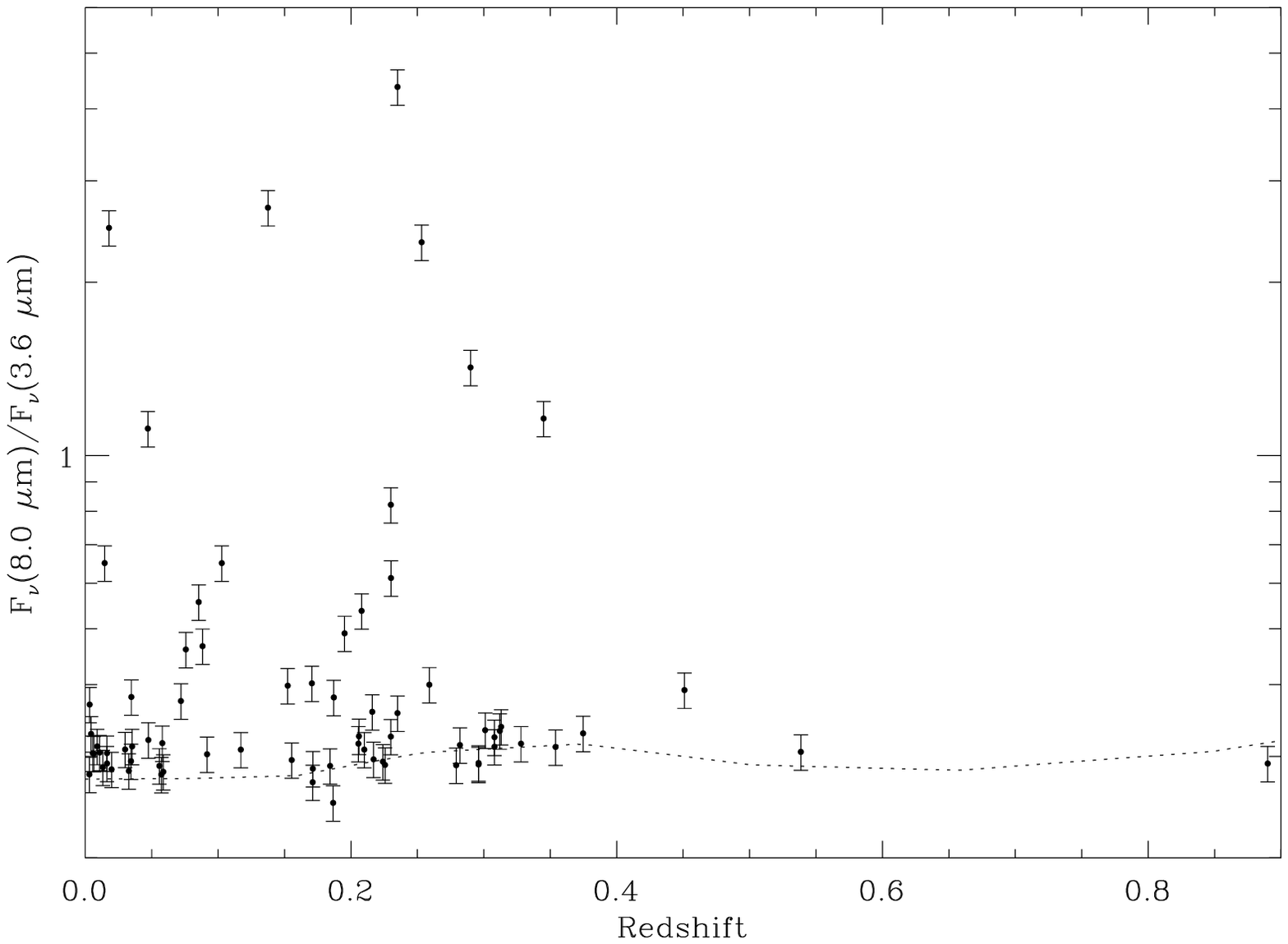}}\quad
\subfigure{\includegraphics[scale=0.4]{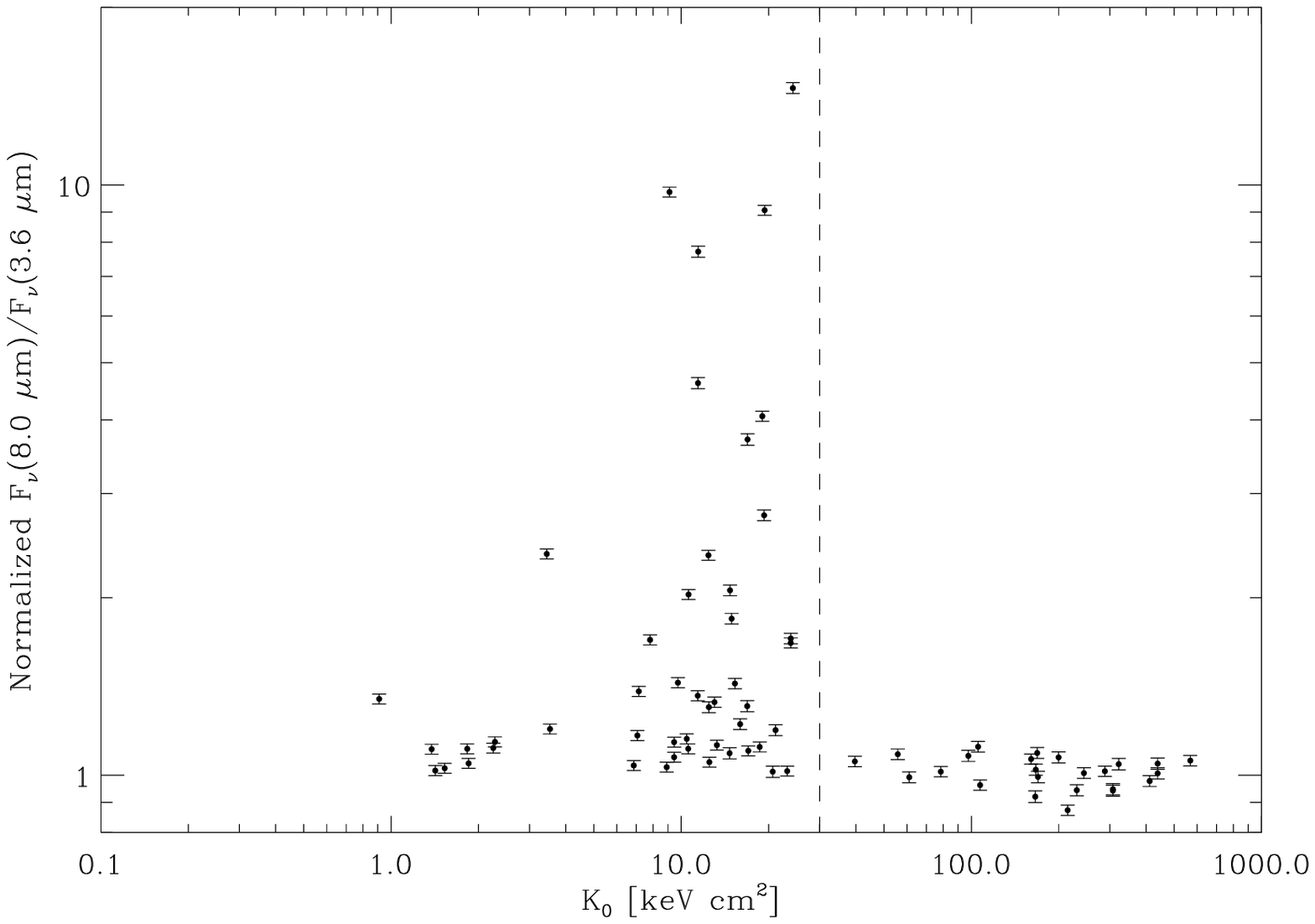}} }
\caption{Redshift dependence of 8.0 $\rm{\mu m}$ to 3.6 $\rm{\mu m}$ ratio. The dotted line represents the expected flux ratio for passively evolving stellar population that is 10 Gyr at $z=0$. 
While IRS spectra of H$\alpha$-emitting BCGs show PAH features that would fall in the 8.0 micron bandpass \cite[e.g.][]{2011ApJ...732...40D} the observed IRAC 8.0 micron color is sensitive to only strong PAH features.
On the right, the flux ratio has been normalized by the passive evolution model and is plotted against the central entropy of the cluster. 
The dotted line identifies the threshold 30~$\rm{keV~cm^2}$. There appears to be a deficit of excess-IR emitters 
in the low $\rm{K_0}$ clusters, but this deficit is likely to be a selection effect since low $K_0$ can only be resolved in the most nearby groups. }
\label{fig:8036}
\end{figure}

\begin{figure}
\centering
\mbox{\subfigure{\includegraphics[scale = .4]{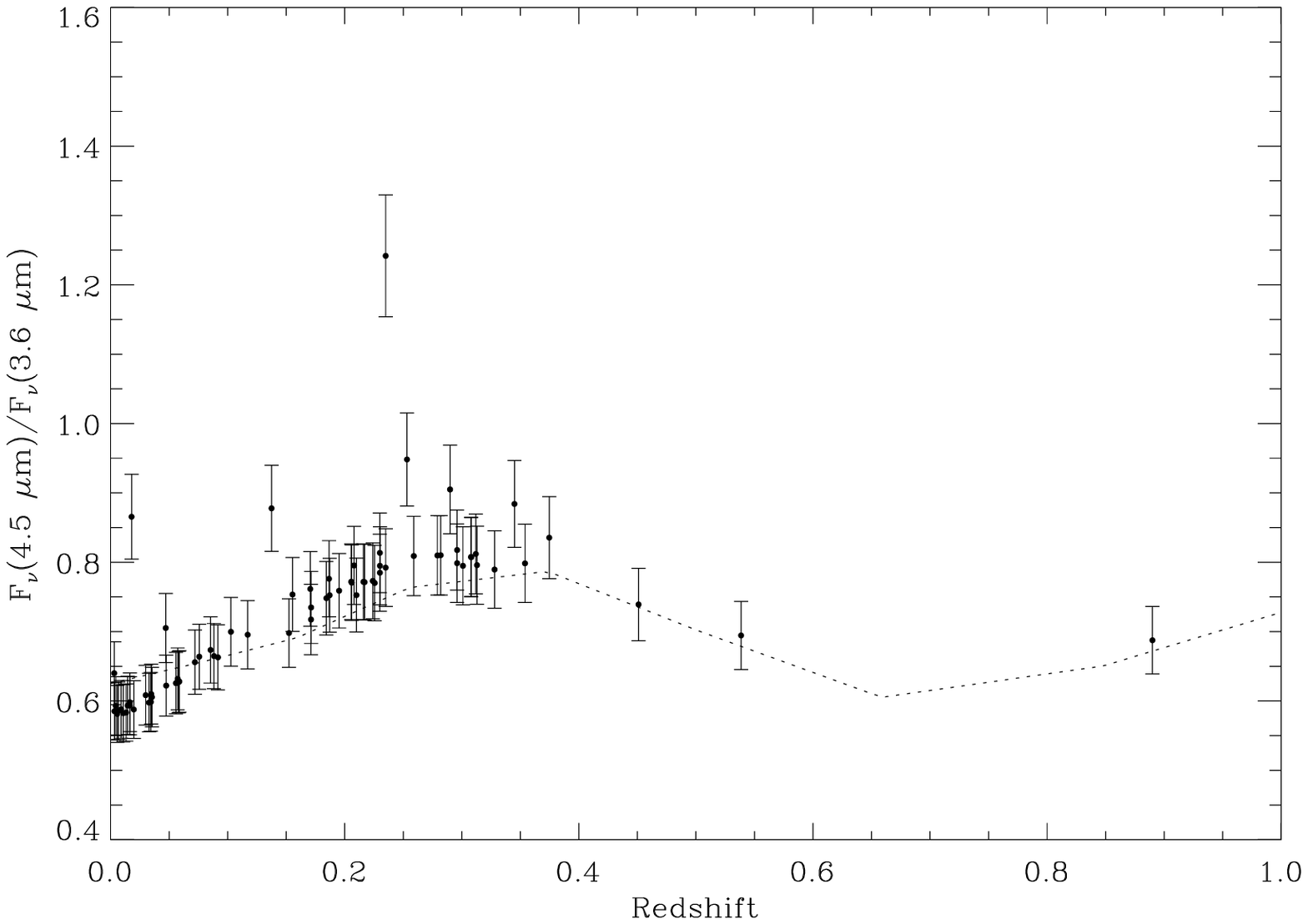}}\quad
\subfigure{\includegraphics[scale=0.4]{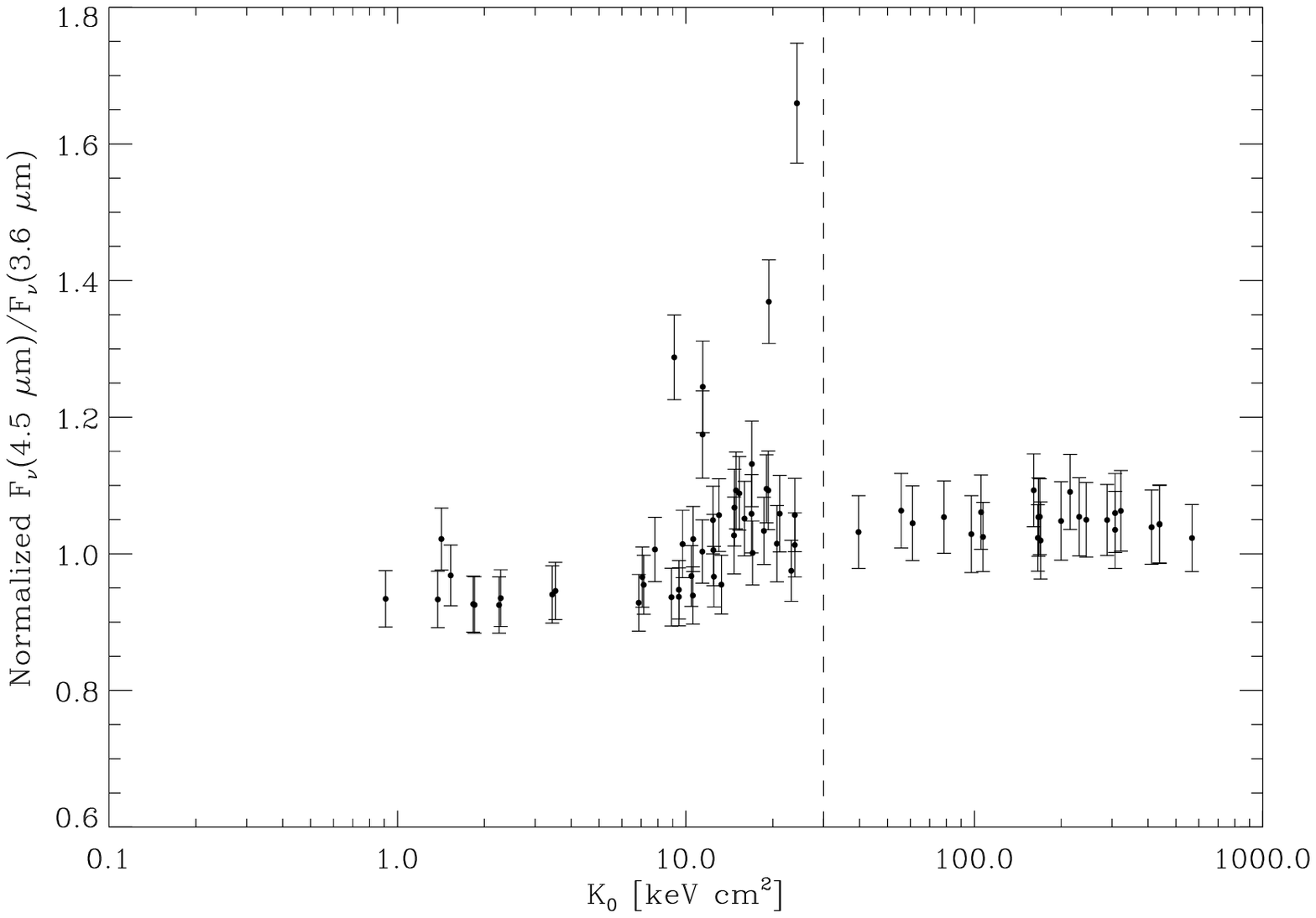}} }
\caption{The left hand figure shows the flux ratio between the 4.5 $\rm{\mu m}$ and the 3.6 $\rm{\mu m}$ flux is plotted as a function of redshift. The dotted line indicates a Starburst99 model for a passively evolving elliptical galaxy with a primarily old stellar population dominated by red giants, with an age of about 10 Gyr at $z=0$. For most of the BCGs the IRAC 4.5 $\rm{\mu m}$ to 3.6 $\rm{\mu m}$ colors are consistent with those of a passively evolving population. In the figure on the right, 
the flux ratio has been normalized by the passive evolution model and is plotted against the central entropy of the cluster. The dotted line again identifies $K_0= 30~\rm{keV~cm^2}$. It is interesting that the handful of BCGs (Abell 426, Abell 1068, Abell 1835, and ZwCl 0857.9+2107) with large excess 4.5 micron emission are located only in clusters with $K_0$ less than the threshold.}
\label{fig:4536}
\end{figure}

\begin{figure}
\includegraphics[scale = .8]{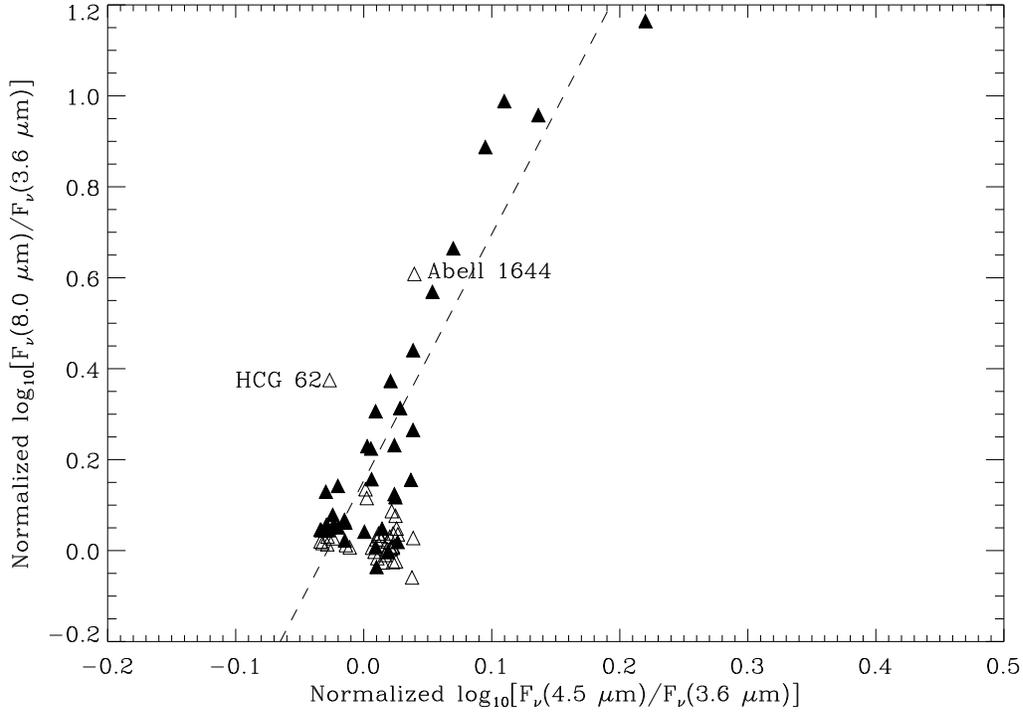}
\caption{8.0 micron to 3.6 micron ratio and 4.5 micron to 3.6 micron ratio. Both ratios have been normalized for passive evolution. The 8.0/3.6 and the 4.5/3.6 ratios are strongly correlated ($r=0.92$(15$\sigma$) for objects with mid-IR detections and/or NUV-K excesses (shown as filled in triangles), which is expected if the excess 4.5 micron emission is generated by processes related to that producing the 8.0 micron emission. Dashed line is a fit to the data; see text. Abell 1644 and HCG 62 do not have blue NUV-K colors, HCG 62 is a 70 micron upper limit, and Abell 1644 wasn't observed by MIPS. Since these ratios use only IRAC data, the uncertainties in the absolute flux calibration are not included. As long as the IRAC calibration was consistent over time, these are precise relative flux ratios. The absolute flux ratios are accurate to about 2\%.}
\label{fig:normalized}
\end{figure}

\begin{figure}
\includegraphics[scale = .8]{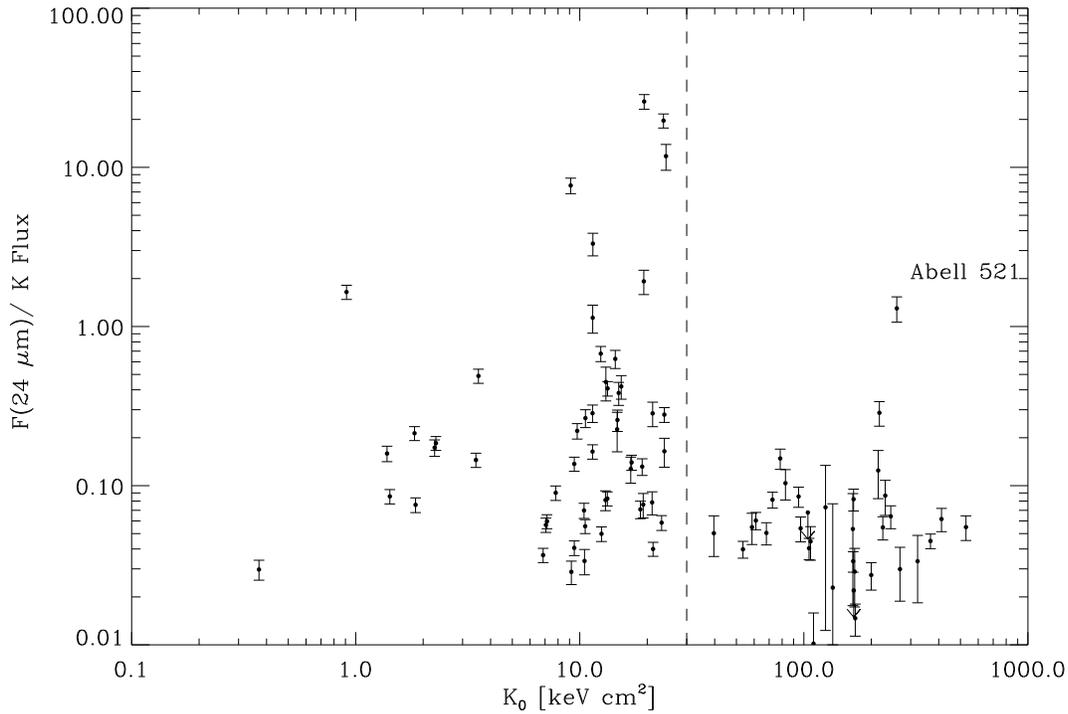}
\caption{24 micron flux to K band flux ratio with central entropy of the cluster. 
BCGs with excess 24 micron flux inhabit clusters with low $K_0$, with the exception of Abell 521.
Even though it is a high $\rm{K_0}$ cluster, the BCG in Abell 521 is in a low entropy, compact, X-ray corona (i.e. a ``mini-cooling core'') which can be associated with BCGs with radio sources and star-formation activity, like BCGs in low $\rm{K_0}$ clusters of galaxies \citep{2006A&A...446..417F}. }
\label{fig:24K}
\end{figure}

\begin{figure}
\includegraphics[scale = .8]{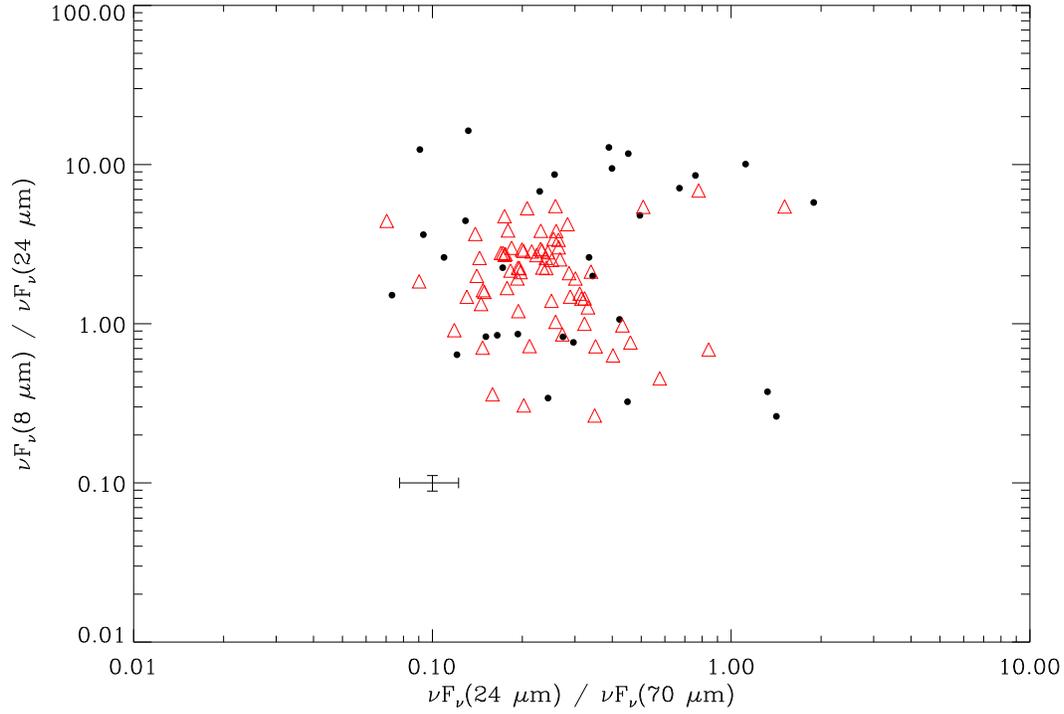}
\caption{Mid IR flux ratio comparison with SINGS galaxies.
This plot is similar to Figure 1 in \citet{2007ApJS..173..377J}. The black dots are our BCGs and the SINGS galaxies \citep{2003PASP..115..928K} are overplotted as red triangles. Ratios for objects with MIPS upper limits at 24 or 70 microns are not plotted but are consistent with the distribution of the detected galaxies. An error bar, representing the standard IRAC and MIPS systematic errors, is plotted to represent a typical error bar. Some nearby BCGs have slightly higher 8.0/24 micron flux ratios than SINGS galaxies, but similar 
24/70 micron flux ratios.}
\label{fig:john1}
\end{figure}

\begin{figure}
\includegraphics[scale = .8]{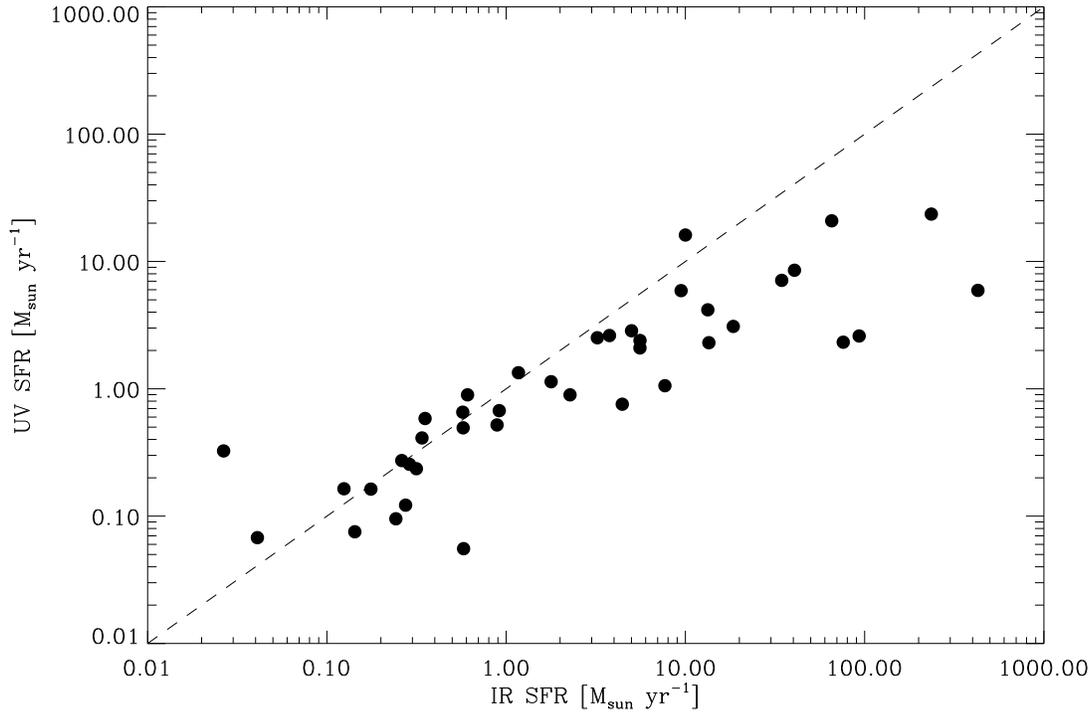}
\caption{UV and IR SFR. The dotted line represents a line of unity. The UV SFR assumes a constant rate of star formation. The model-derived IR star formation rates are consistent with star formation rates measured with a MIPS 70 $\mu$m SFR estimate as shown in Figure~\ref{fig:SFRcomp}. Those objects which fall below the line, BCGs with excess IR star formation, are similar to starburst galaxies, in the sense that for the most luminous
star-forming galaxies, most of the star-formation is obscured.}
\label{fig:SFR}
\end{figure}

\begin{figure}
\centering
\mbox{\subfigure{\includegraphics[scale = .4]{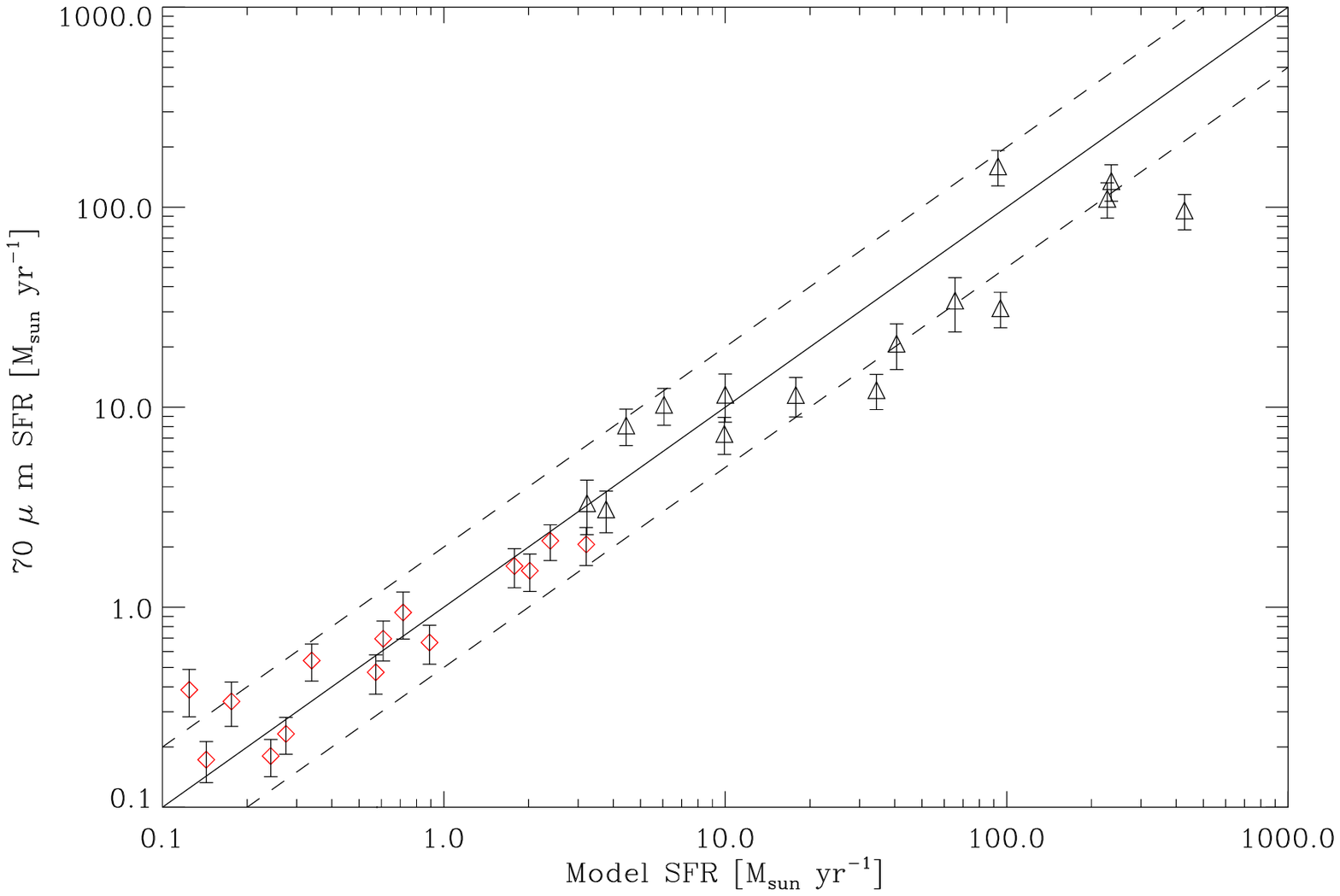}}\quad
\subfigure{\includegraphics[scale=0.4]{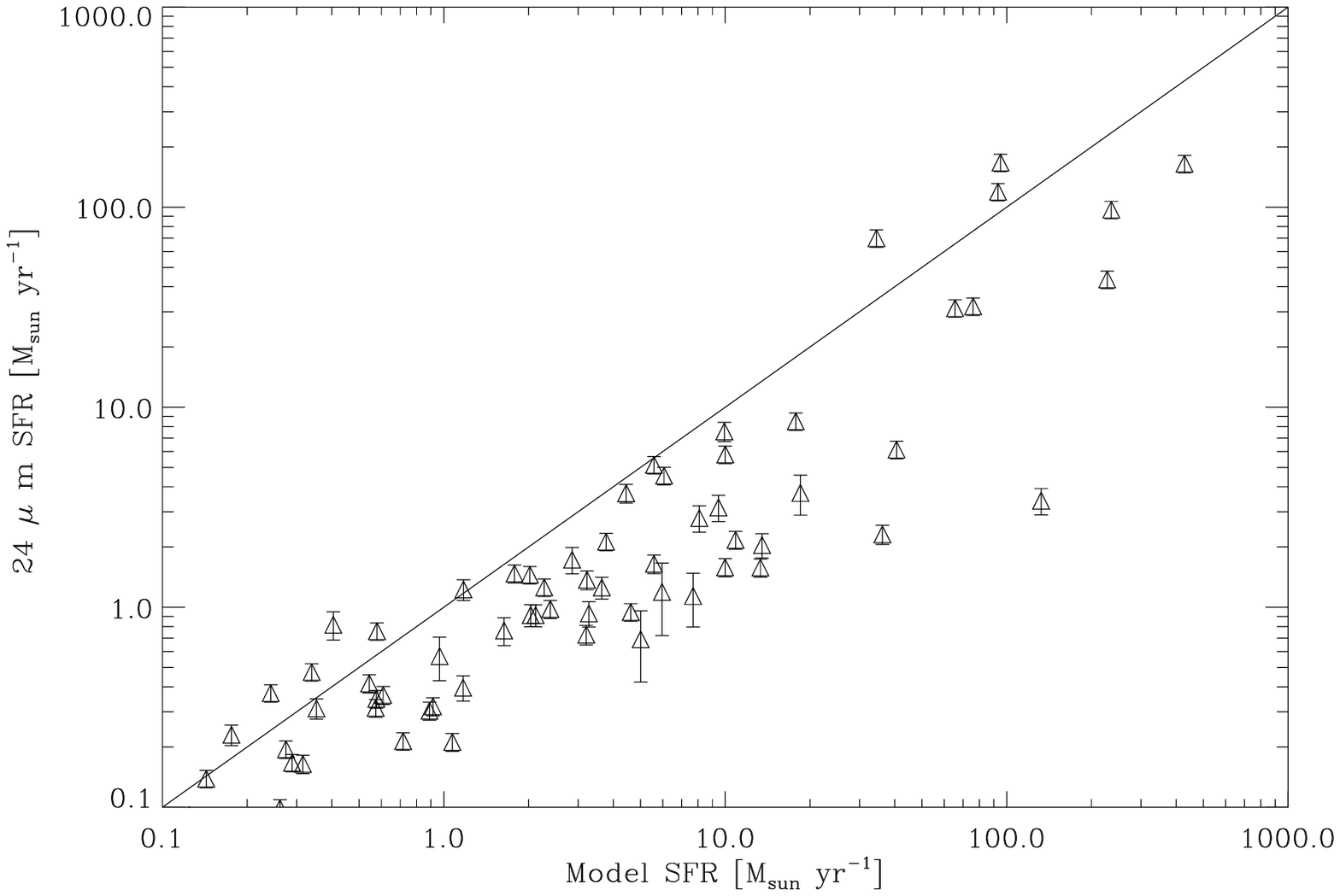}} }
\caption{Comparison of model IR SFR to single band IR SFR. We compare the estimates from the Groves model star formation rates to the star formation rate estimates using the 70 micron luminosity in the left plot. The line represents a line of unity, not a fit. The dotted lines represent the boundary for a difference of a factor of two in star formation rate. The black triangles represent the high luminosity relation given in \citet{2010ApJ...714.1256C} while the red diamonds use their relation for galaxies with low IR luminosities (and therefore a larger amount of the IR flux is produced by dust heated by evolved stars rather than hot stars). The right plot is a similar plot relating the 24 micron luminosity to the Groves model star formation rates. The 24 micron SFRs tend to have lower estimates as the 70 micron luminosities are found nearer to the peak of the dust blackbody and therefore more representative of the total IR luminosity and SFR. For the most luminous 70 microns galaxies, it appears that the model fit tends to overpredict the SFR and luminosity relative to the 70 micron flux estimate.}
\label{fig:SFRcomp}
\end{figure}

\begin{figure}
\includegraphics[scale = .8]{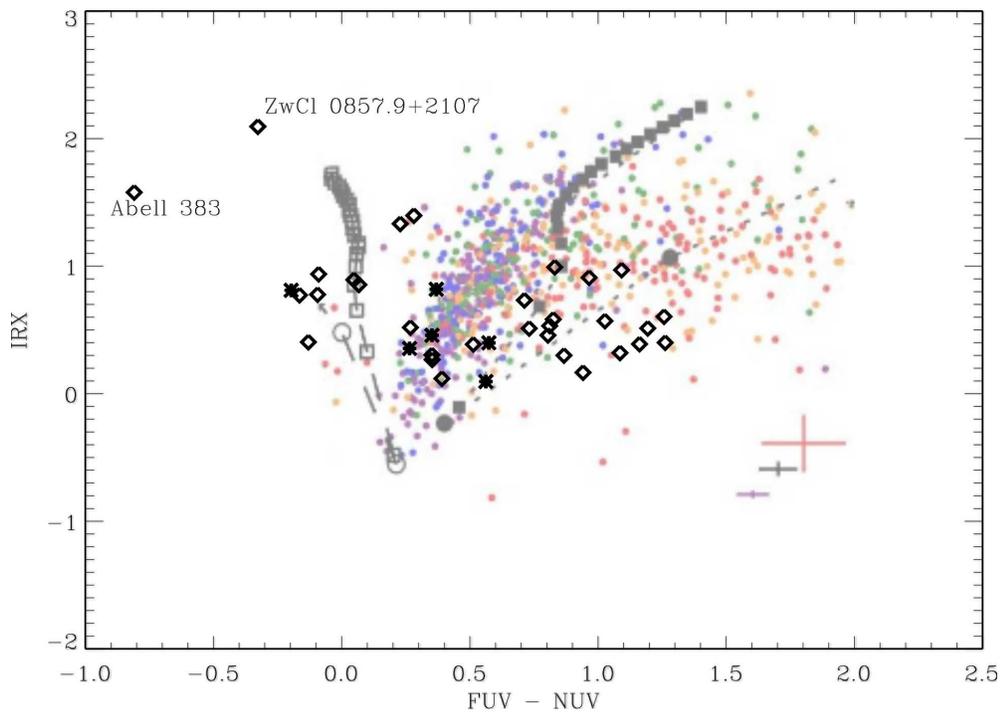}
\caption{IR excess and UV color. The IR excess (IRX) is defined in \citet{2007ApJS..173..377J} to be the ratio of IR to UV luminosity. Objects from the cool core sample of \citet{2010ApJ...719.1844H} are marked with X's. Figure 6a from \citet{2007ApJS..173..377J} is plotted in the background on this graph. }
\label{fig:john6}
\end{figure}

\begin{figure}
\includegraphics[scale = .8]{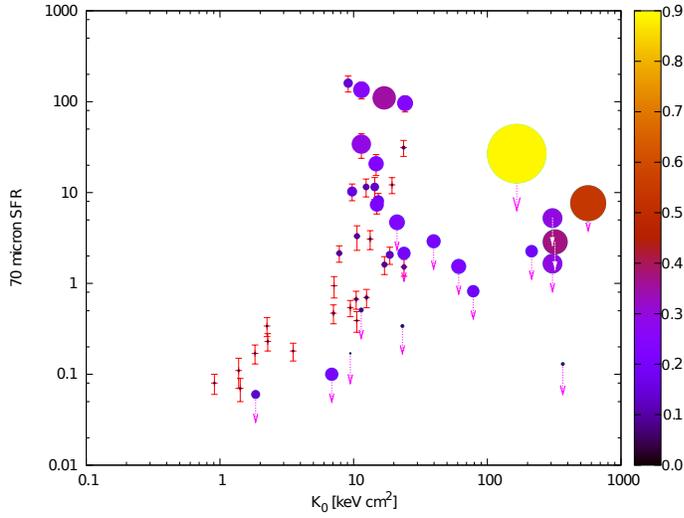}
\caption{Relation between 70 micron SFR and central entropy. The objects are color coded and sized based on redshift (i.e. higher redshift, larger size). The color code ranges from redshift of 0.0 to 0.9. The ``trend'' that is seen in the lowest central entropy clusters is not a physical trend, but it is the same selection effect noted in Figure~\ref{fig:NUV}. Note that all BCGs in clusters with $K_0>30~\rm{keV~cm^2}$ in this plot have only upper limits.}
\label{fig:k0_SFR}
\end{figure}